\documentclass[preprint,12pt,authoryear,review]{elsarticle}
\usepackage[margin=2cm]{geometry}
\usepackage[margin=5pt,font=small,labelfont=bf]{caption}
\usepackage{graphicx,amsmath,amssymb,fge,setspace,xspace,color}
\usepackage[displaymath,mathlines]{lineno}
\definecolor{darkblue}{rgb}{0.0,0.0,0.6}
\usepackage[colorlinks=true,breaklinks=true,linkcolor=darkblue,
            urlcolor=darkblue,anchorcolor=darkblue,citecolor=darkblue]{hyperref}
\usepackage{multirow} \usepackage{pdflscape}

\journal{Earth Planet Sci Letters}

\newcommand{\T}{\rule{0pt}{3.2ex}} % Top strut 
\newcommand{\B}{\rule[-1.2ex]{0pt}{0pt}} % Bottom strut

\newcommand{\textsup}[1]{\textsuperscript{#1}}

\newcommand{\infd}{\textrm{d}}

\newcommand{\vs}{\mathbf{v}_s}
\newcommand{\vl}{\mathbf{v}_{\ell}}

\newcommand{\phieq}{\phi^\mathrm{eq}}
\newcommand{\phis}{(1-\phi)}

\newcommand{\q}{\mathbf{q}}

\newcommand{\Deltarho}{\Delta \rho}

\newcommand{\cbar}{\bar{c}^i}

\newcommand{\csi}{c_s^i}
\newcommand{\cli}{c_{\ell}^i}
\newcommand{\cseq}{c_s^{i,\mathrm{eq}}}
\newcommand{\cleq}{c_{\ell}^{i,\mathrm{eq}}}

\newcommand{\R}{\mathcal{R}}

\newcommand{\Ki}{K^i}

\newcommand{\G}{\Gamma}
\newcommand{\Gi}{\Gamma^i}

\newcommand{\RG}{\mathcal{R}_{\Gamma}}

\newcommand{\degC}{$^\circ$C}
\newcommand{\water}{H$_2$O}
\newcommand{\carbon}{CO$_2$}
\newcommand{\e}[1]{$\times10^{#1}$}
\newcommand{\qb}{q_\mathrm{base}^i}
\newcommand{\qm}{q_\mathrm{melt}^i}
\newcommand{\qf}{q_\mathrm{focus}^i}
\newcommand{\Rm}{R_\mathrm{melt}^i}
\newcommand{\Rf}{R_\mathrm{focus}^i}

\begin{document}

\begin{frontmatter}

  %% TITLE, AUTHORS & AFFILIATIONS

  \title{Volatiles beneath mid-ocean ridges: deep melting, channelised transport, focusing, and metasomatism}
  \author[oxf,stf]{Tobias Keller\corref{cor}}
  \ead{tobias.keller@stanford.edu} 	
  \author[oxf]{Richard F. Katz}
  \author[min]{Marc M. Hirschmann}

  \address[oxf]{University of Oxford, Oxford, UK.}
  \address[stf]{Stanford University, Stanford CA, USA.} 
  \address[min]{University of Minnesota, Minneapolis MN, USA.}
  \cortext[cor]{Corresponding author}

  %% ABSTRACT A concise and factual abstract is required. The abstract
  %% should state briefly the purpose of the research, the principal
  %% results and major conclusions. An abstract is often presented
  %% separately from the article, so it must be able to stand
  %% alone. For this reason, References should be avoided, but if
  %% essential, then cite the author(s) and year(s). Also,
  %% non-standard or uncommon abbreviations should be avoided, but if
  %% essential they must be defined at their first mention in the
  %% abstract itself.

  %% GRAPHICAL ABSTRACT (??)  Although a graphical abstract is
  %% optional, its use is encouraged as it draws more attention to the
  %% online article. The graphical abstract should summarize the
  %% contents of the article in a concise, pictorial form designed to
  %% capture the attention of a wide readership. Graphical abstracts
  %% should be submitted as a separate file in the online submission
  %% system. Image size: Please provide an image with a minimum of 531
  %% × 1328 pixels (h × w) or proportionally more. The image should be
  %% readable at a size of 5 × 13 cm using a regular screen resolution
  %% of 96 dpi. Preferred file types: TIFF, EPS, PDF or MS Office
  %% files. See https://www.elsevier.com/graphicalabstracts for
  %% examples.

\begin{abstract}
  Deep-Earth volatile cycles couple the mantle with near-surface
  reservoirs.  Volatiles are emitted by volcanism and, in particular,
  from mid-ocean ridges, which are the most prolific source of
  basaltic volcanism.  Estimates of volatile extraction from the
  asthenosphere beneath ridges typically rely on measurements of
  undegassed lavas combined with simple petrogenetic models of the
  mean degree of melting.  Estimated volatile fluxes have large
  uncertainties; this is partly due to a poor understanding of how
  volatiles are transported by magma in the asthenosphere.  Here, we 
  assess the fate of mantle volatiles through numerical simulations of 
  melting and melt transport at mid-ocean ridges.  Our simulations are 
  based on two-phase, magma/mantle dynamics theory coupled to an 
  idealised thermodynamic model of mantle melting in the presence of 
  water and carbon dioxide.  We combine simulation results with 
  catalogued observations of all ridge segments to estimate a range of 
  likely volatile output from the global mid-ocean ridge system.  We 
  thus predict global MOR crust production of 66--73~Gt/yr 
  (22--24~km\textsup{3}/yr) and global volatile output of 
  52--110~Mt/yr, corresponding to mantle volatile contents of 
  100--200~ppm.  We find that volatile extraction is limited: up to 
  half of deep, volatile-rich melt is not focused to the axis but is 
  rather deposited along the  LAB.  As these distal melts crystallise 
  and fractionate, they metasomatise the base of the lithosphere, 
  creating rheological heterogeneity that could contribute to the 
  seismic signature of the LAB.
\end{abstract}

%% KEYWORDS Immediately after the abstract, provide a maximum of 6
%% keywords, using American spelling and avoiding general and plural
%% terms and multiple concepts (avoid, for example, 'and', 'of'). Be
%% sparing with abbreviations: only abbreviations firmly established
%% in the field may be eligible. These keywords will be used for
%% indexing purposes.

\begin{keyword}
  Mid-ocean ridge degassing \sep mantle melting \sep reactive channels 
  \sep melt focusing \sep deep volatile cycles 
  \sep magma/mantle dynamics
\end{keyword}

\end{frontmatter}
\newpage

%% HIGHLIGHTS Highlights are mandatory for this journal. They consist
%% of a short collection of bullet points that convey the core
%% findings of the article and should be submitted in a separate
%% editable file in the online submission system. Please use
%% 'Highlights' in the file name and include 3 to 5 bullet points
%% (maximum 85 characters, including spaces, per bullet point). See
%% https://www.elsevier.com/highlights for examples.

%\section*{Highlights}
%\begin{itemize}
%\item Deep, water- and carbon dioxide-rich melting causes channelized
%  magma transport in 2D computational simulations of mid-ocean ridge 
%  magmatism.
%\item Magma focusing towards ridge axis controlled by compaction
%  length and width of volatile-free melting regime, independent of 
%  volatiles.
%\item Volatile partitioning between ridge extraction and
%  lithosphere metasomatism controlled by melt focusing regime.
%\item Volatile metasomatism along LAB leaves lower lithosphere
%  significantly enriched in volatiles, creating weak layer.
%\item Global estimates for MOR crust production and volatile
%  extraction rates inferred from computational parameter study.
%\end{itemize}
%\newpage

\tableofcontents

\newpage
%\linenumbers

%% PAPER LENGTH EPSL has a restricted article length of not more than
%% 6500 words in the main text, excluding the abstract, figure
%% captions and references, and a total number of figures and tables
%% not to exceed 10 (where both are counted together). Large tables
%% should be submitted as part of Supplementary Material.

%% MAIN BODY

%% INTRODUCTION State the objectives of the work and provide an
%% adequate background, avoiding a detailed literature survey or a
%% summary of the results.

\section{Introduction \label{sect:Introduction}}
%% SCIENCE QUESTIONS 1. What controls focusing of melt and volatiles
%% to ridge axis?  2. What controls volatile extraction fluxes at
%% ridge axis?  3. What are implications for global MOR extraction
%% estimates?  4. What are implications for the compositional and
%% rheological state of the oceanic LAB?

\subsection{Deep volatile cycles}
Deep-Earth volatile cycles are a result of fluxes between
near-surface, shallow-mantle and deep-mantle reservoirs.  These fluxes
are associated with transport processes in the solid Earth that may be
modified by the volatiles themselves \citep[e.g.,][]{hirth96,
  dasgupta10}.  Volatile cycles impact aspects of planetary dynamics
including mantle convection and partial melting.  Partial melting is
the key process by which volatiles are released from the mantle to
near-surface reservoirs; it influences deep and shallow phenomena
including the formation of the oceanic \citep{hirth96, asimow04} and
continental crust \citep{muntener01, ulmer01}, the long-term tectonic
regime of a planet \citep{mian90, regenauerlieb01} and, potentially,
variations of the surface climate \citep{huybers09, crowley14, burley15, 
tolstoy15, huybers17}.  The thermodynamic effect of mantle
volatiles on partial melting and, in particular, that of water and
carbon dioxide, is now well documented \citep[e.g.,][]{hirschmann99,
  asimow03, dasgupta13}. A key problem, however, remains almost
unexplored: the dynamics of melt transport in the presence of
volatiles. A theoretical framework to investigate these dynamics has
recently been developed by \cite{keller16}.

Volatile-enriched partial melting and melt transport is relevant to
all environments in which partial melting occurs; perhaps most of all
to subduction zones, where melt production is primarily driven by the
input of volatiles.  In this manuscript, however, we focus on
divergent plate margins, which are a simpler geodynamic context. The
mid-ocean ridge (MOR) system is the primary avenue for volatile
extraction from the mantle \citep{resing04, dasgupta10, kelemen15}. A
simple estimate of the volatile output from ridges is the product of 
the formation rate of oceanic crust ($\sim24$~km\textsup{3}/yr or
72~Gt/yr, \cite{crisp84}) and the mean volatile concentration in 
undegassed primary mid-ocean-ridge basalt (MORB) \citep[e.g.,][]
{michael15, rosenthal15}. \cite{resing04} constrain the MOR \carbon{} 
emission rate to a range of 22--88~Mt~\carbon/yr, \cite{dasgupta10} 
produce estimates of 46--324, \cite{cartigny08} of 44--220, and 
\cite{kelemen15} of 29--154~Mt~\carbon/yr. Some of these include 
contributions from oceanic intraplate volcanism.

Such estimates are the basis for currently accepted volatile output
from ridges, but bypass important
considerations.  For example, relating fluxes to the concentration of
volatiles in the sub-ridge mantle requires constraints on the locus of
partial melting in the mantle as well as on the efficiency of volatile
transport with ascending melts.  Also, simple budget calculations
cannot constrain the fraction of volatile-rich melts that are emplaced
into the oceanic lithosphere rather than focused to the ridge axis.
Any such emplaced melts may influence the properties of the oceanic
lithosphere and contribute to volatile release during subduction.
Moreover, mean flux estimates do not allow exploration of dynamic
controls such as viscosity and permeability \citep{braun00, kono14},
channelised flow \citep{kelemen95a}, and spatial variations in mantle
potential temperature and composition \citep{dalton14}.

\subsection{Volatiles in mid-ocean ridge magmatism}
Small concentrations of volatiles greatly expand the volume of mantle
that produces partial melt beneath ridges by causing melting to
initiate at greater depths. For typical sub-ridge potential
temperatures, volatile-free melting may occur above 60--75~km, but as
little as 100~ppm \water{} depresses the onset of melting to
100--120~km depth \citep{hirth96, hirschmann99}. Comparable amounts of
\carbon{} lead to an onset of melting between 150 and 300~km depth,
depending on the redox state of the mantle \citep{dasgupta10}.
Moreover, due to the widening of the upwelling region with depth
\citep{lachenbruch76, spiegelman87}, deepening the melting regime
produces small fractions of melt at considerable lateral
distance away from the axis. These distal, low-degree melts are rich
in incompatible trace elements as well as in volatiles. The proportion
of distal melts that are focused to the ridge axis rather than frozen
into growing oceanic lithosphere or erupted at off-axis seamounts is
uncertain \citep{galer86, plank92}.  Furthermore, erupted MORB has
long been considered a mixture of polybaric melts, the composition of 
which is sensitive to the shape of the melting regime 
\citep{ohara85, plank92, asimow03}. Lastly, the consequences of 
rheological weakening by volatiles are not yet well understood
\citep{hirth96, braun00}. Resolving the dynamic aspects of MORB 
petrogenesis therefore requires an improved understanding of the 
reactive segregation of deep, volatile-rich melts.

Such deep melting produces melt fractions much less than 1\%
\citep[e.g.,][]{plank92, hirth96, asimow04, dasgupta06}.  Evidence
from short-lived radiogenic products of uranium decay ($^{230}$Th,
$^{231}$Pa, and $^{226}$Ra) indicates that very small melt fractions
segregate from their deep sources and ascend rapidly to the surface
\citep{spiegelman93d, mckenzie00, elliott03}.  This may be facilitated
by high-permeability melt channels \citep{aharonov95, spiegelman01}. 
\cite{keller16} showed that such channels may form when deep, 
volatile-rich melt fluxes the primary basaltic melting regime from 
below.  Quantifying the consequences of incipient melting, channelised 
melt transport, and magmatic focusing on MOR volatile output is the 
goal of this study.

\subsection{Method summary}
We use two-dimensional numerical simulations of two-phase,
multi-component, reactive flow in the asthenosphere to model magma
genesis and transport beneath a MOR. The model is calibrated to
reproduce accepted features of melting in an upwelling mantle
containing \water{} and \carbon{} in low concentrations.  Crustal
thickness is used as an observational constraint on MOR models
\citep{white01}; we generate predictions for comparison with data by
computing the rate at which magma is delivered to the ridge axis. To
scale our model results from individual instances to the ensemble of
ridge segments that comprise the global MOR system, we use a catalogue
of kinematic, chemical, and thermal parameters compiled by
\cite{gale14} and \cite{dalton14}. Empirical scaling laws fitted to 
the output of dynamic simulations are combined with this catalogued 
data to create estimates of crust production and volatile extraction 
for the global MOR system.

Section~\ref{sect:model} introduces the computational model used to
generate simulations and discusses the leading-order features and
parametric controls of their output.  Section~\ref{sect:fluxes}
discusses melt focusing, crust production and volatile extraction
predicted by simulations; section~\ref{sect:rates} develops estimates
of global MOR volatile output; metasomatism of the oceanic lithosphere
is considered in section~\ref{sect:lab}. 
Section~\ref{sect:conclusions} summarises our findings and offers some
conclusions.

% Previous models of MORB petrogenesis have reduced partial melting and 
% melt focusing to a mixing problem based on analysis of the residual 
% mantle column~(RMC) --- the result of steady-state melt extraction 
% across the melting regime \citep[e.g.,][]{klein87, mckenzie88, 
% plank92, asimow03}.

%% METHOD Provide sufficient detail to allow the work to be
%% reproduced. Methods already published should be indicated by a
%% reference: only relevant modifications should be described.

\section{Mid-ocean ridge simulations \label{sect:model}}

\subsection{Theory, parameters, and numerical solutions}
Simulations are based on numerical solutions to a system of partial
differential equations comprising statements of mass, momentum, 
and energy conservation for two phases (liquid and solid), and four 
thermochemical components. Solutions are computed on a two-dimensional 
domain representing half of a symmetrical spreading center. The 
domain extends to a depth of 200~km and a width of 300~km from the  
axis. Model derivation and numerical implementation are given in 
\cite{keller16}; the theory is built upon that of \cite{mckenzie84} 
and \cite{rudge11}. 

Magma and rock compositions are modelled in a four-component
compositional space of DUN $+$ MORB $+$ hMORB $+$ cMORB. These
chemical components represent the residue of mantle melting (dunite),
the product of volatile-free decompression melting (basalt), and the
products of hydrated and carbonated melting at depth (hydrated,
carbonated basalt), respectively. The latter two components contain
volatiles at fixed concentrations of 5~wt\%~\water{} in hMORB, and
20~wt\%~\carbon{} in cMORB. Linear kinetic reaction rates are
calculated with the R\_DMC method \citep{keller16, rdmc-repo}, using a 
constant rate factor of $\R = 3$~kg/m\textsup{3}/yr (corresponding to 
a reaction time of $\sim$1~kyr). Density is taken as constant and 
uniform for both phases, except in body-force terms where a density 
difference between melt and solid of $\Deltarho = 500$~kg/m\textsup{3} 
drives segregation. See Appendix \ref{sect:thermodyn} for a summary of the R\_DMC model for thermodynamic equilibrium and melt-rock reactions.

The viscosity of mantle rock is reduced by water and 
melt \citep{hirth96, mei02}, and volatile-bearing silicate melt is 
weakened by both \water{} and \carbon{}. Permeability 
depends on grain size and melt fraction according to the Kozeny-Carman 
relation. For details of constitutive laws see Appendix 
\ref{sect:rheology}.

Plate spreading at a half-rate $u_0$ is imposed along the top boundary
of the model domain; the bottom and off-axis side boundaries are open
to inflow and outflow of the mantle in passive response to plate
spreading. Symmetry conditions are applied along the vertical boundary
beneath the ridge axis. Melt extraction at the axis occurs through an
imposed patch of $4\times8$~km size where melting and
freezing reactions are disallowed. For numerical convenience, a 
maximum solidus depth of 160~km is imposed to keep the lower boundary 
of the domain melt-free at all times.

Calculations are initialised with the temperature solution to the
plate-cooling half-space problem. Some pre-depletion of the mantle
composition is imposed in the melting region. The initial temperature
is then limited point-wise by the solidus temperature such that the
melt fraction is initially zero. These arrangements minimise the time
required for model evolution from the initial condition to an
approximately steady state (the ``spin-up'' time) \citep{katz08, 
  katz10}.

Simulations at reference resolution of 1~km per grid cell have a
numerical problem size of 744k degrees of freedom and are computed on
64~cores of the BRUTUS cluster at ETH Zurich, Switzerland.  Numerical
solutions are obtained using a parallel Newton-Krylov method
leveraging the PETSc toolkit \citep{petsc-web-page, petsc-user-ref,
  katz07}.

%% RESULTS Results should be clear and concise.

%% DISCUSSION This should explore the significance of the results of
%% the work, not repeat them. A combined Results and Discussion
%% section is often appropriate. Avoid extensive citations and
%% discussion of published literature.

%% Metrics as function of parameter variations
%% \item[-] Steady-state magnitude of crustal thickness, volatile
%%   extraction fluxes
%% \item[-] Polynomial fitting functions through simulation data as
%%   function of mantle parameters
%% \item[-] Melt focusing distance defined by melt streamlines and
%%   "equivalent focusing distance"
%% \item[-] Melt production and melt focusing ratios: mass flux of
%%   melt produced by mass flux into base of domain; mass flux of melt
%%   extraction by mass flux of melt produced.
%%
%%   Additional observations and calculations
%% \item[-] Global estimates of MOR crustal production and volatile
%%   extraction rates.
%% \item[-] Characteristics of melt transport regime (interaction of
%%   reactive channels, compaction length, focusing distance)
%% \item[-] Rheological and compositional state of LAB away from ridge
%%   axis

\subsection{Reference model}
The reference model simulates an intermediate-spreading ridge with
$u_0=3$~cm/yr, a mantle potential temperature $T_m=1350$~\degC, and a
mantle bulk composition $\bar{c}^i$ (in wt\%) of 74.75~DUN, 25~MORB,
0.2~hMORB, and 0.05~cMORB. The concentrations of the latter two
correspond to volatile contents of 100~wt~ppm each of \water{} and
\carbon{}. A white-noise perturbation field is low-pass filtered to a
minimum wavelength of $\sim$5~km and used to add compositional 
heterogeneity in mantle fertility (MORB content) at a relative 
amplitude of $\pm$10\%.  The identical perturbation is used to 
introduce mantle heterogeneity in all simulations reported here.

\begin{figure}[htb]
  \centering
  \includegraphics[width=\textwidth]{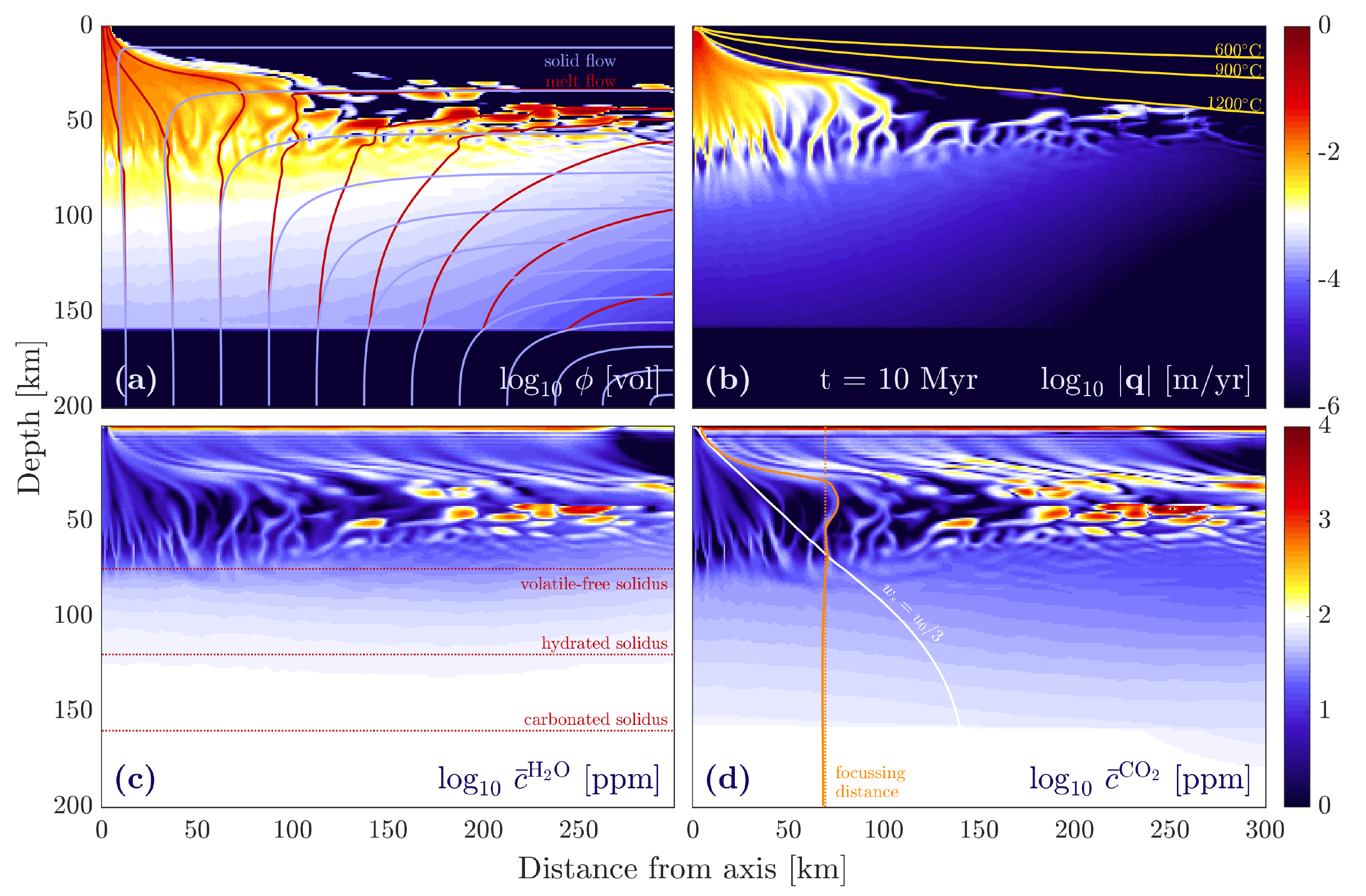}
  \caption{Output from reference simulation at 10~Myr. Panels show 
  \textbf{(a)} melt fraction, with melt (red) and solid (blue) 
  streamlines; \textbf{(b)} Darcy flux magnitude, with 600, 900 and 
  1200~\degC{} isotherms; \textbf{(c)} bulk water concentration, with 
  calibrated solidus depths for volatile-free, hydrated and carbonated 
  mantle; \textbf{(d)} bulk carbon dioxide concentration, with contour 
  of primary upwelling and melt producing domain (white) and melt  
  streamline delimiting the melt focusing domain (orange), with  
  equilibrium focusing distance $x_e$ (dotted) for comparison.}
  \label{fig:1}
\end{figure}

Figure~\ref{fig:1} shows a snapshot of the reference model after
10~Myr.  At this stage, plate spreading and mantle upwelling have swept 
the full width and depth of the domain and thus dynamics have evolved 
well past the effects of initial conditions. The general features of the 
magmatic system are representative of all simulations presented here.  A 
key feature is the coalescing network of magmatic channels focusing melt 
towards the ridge axis.  These channels are most notable across a depth 
range of 50--80~km below the ridge axis and are characterised by 
increased melt fraction (panel~(a)), magnitude of Darcy flux 
$|\q| = \phi|\vl-\vs|$ (panel~(b)), and volatile concentrations 
(panels~(c)~\&~(d)).

The network is formed of reactive-dissolution channels as described in
\cite{keller16}. Channelisation is driven by the corrosivity of deep,
low-degree, volatile-rich melts. These melts flux the base of the
volatile-free melting regime where they create a Reactive Infiltration
Instability \citep{aharonov95, szymczak14}. Enhanced flux of deep melt
leads to dissolution of the MORB component, increased porosity and
permeability, and further enhancement of melt flux. Reactive channels
thus form approximately parallel to the melt flow direction.

\begin{figure}[htb]
  \centering
  \includegraphics[width=\textwidth]{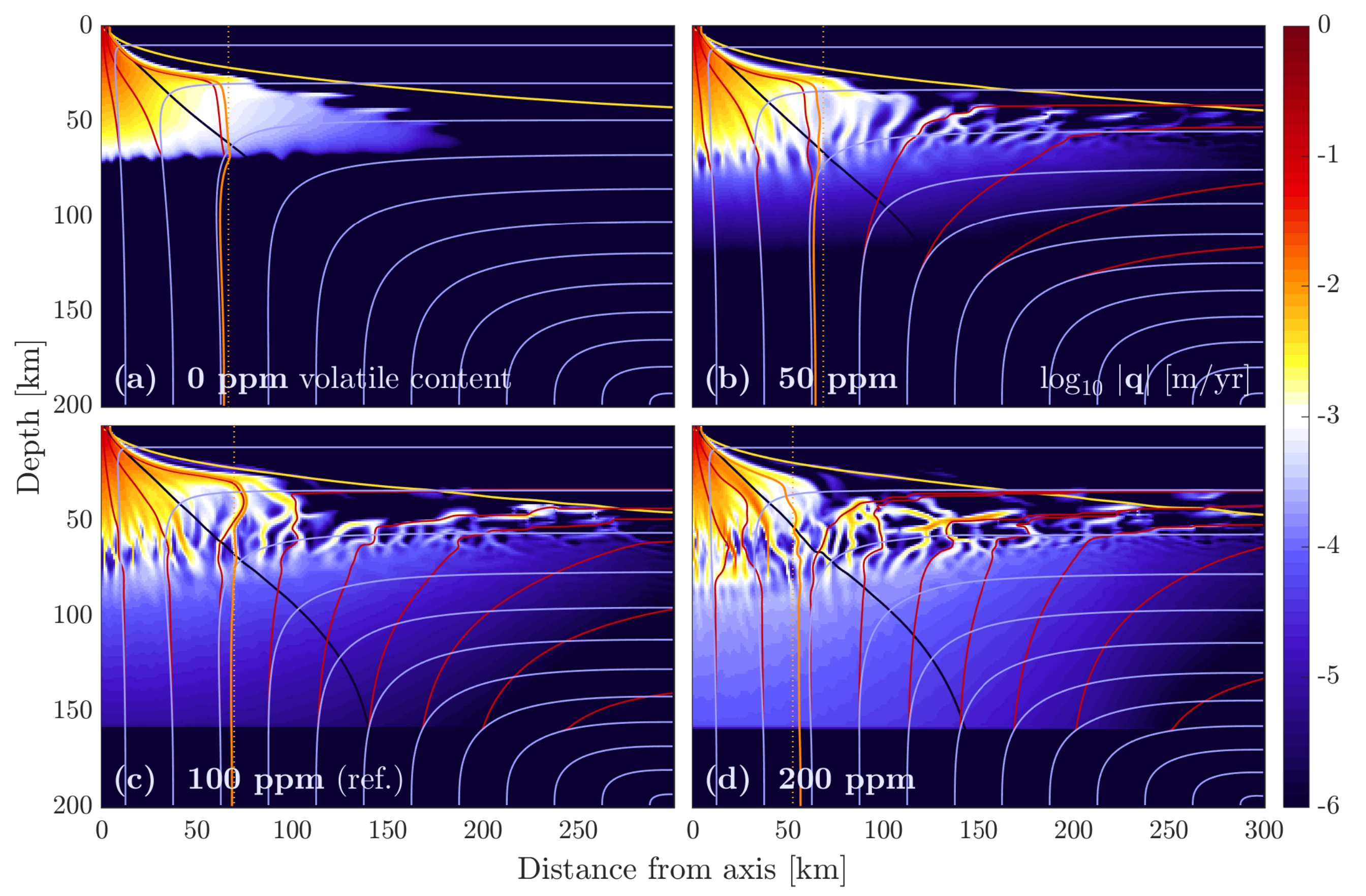}
  \caption{Darcy flux magnitude $\vert\q\vert$ at $t=10$~Myr for
    simulations with \textbf{(a)} no volatiles, \textbf{(b)} 50~ppm,
    \textbf{(c)} 100~ppm (reference), and \textbf{(d)} 200~ppm each
    \water{} and \carbon{} in the mantle source. Streamlines for melt
    (red) and solid (blue) flow, outline of primary upwelling domain
    (black), and melt focusing domain (orange), with $x_e$
    (dotted) for comparison.}
  \label{fig:2}
\end{figure}

Figure~\ref{fig:2} shows a comparison of the Darcy flux in simulations
with volatile contents of 0, 50, 100 (ref.~case), and 200~ppm. The
distributed flow regime in the volatile-free case of panel~(a)
indicates that the perturbation of the compositional field does not
itself cause channelisation. However, with increasing volatile content
(panels~(b)--(d)) the degree of flow localisation increases. This
trend reinforces the argument that the presence of volatile elements
in the mantle source causes channelisation.  In our calculations, as
little as 50~ppm each of \water{} and \carbon{} is sufficient to
initiate channelling. This value is at the low end of estimates for
the MORB-source mantle.

\subsection{Volatile extraction by melt focusing}
\begin{figure}[htb]
  \centering
  \includegraphics[width=\textwidth]{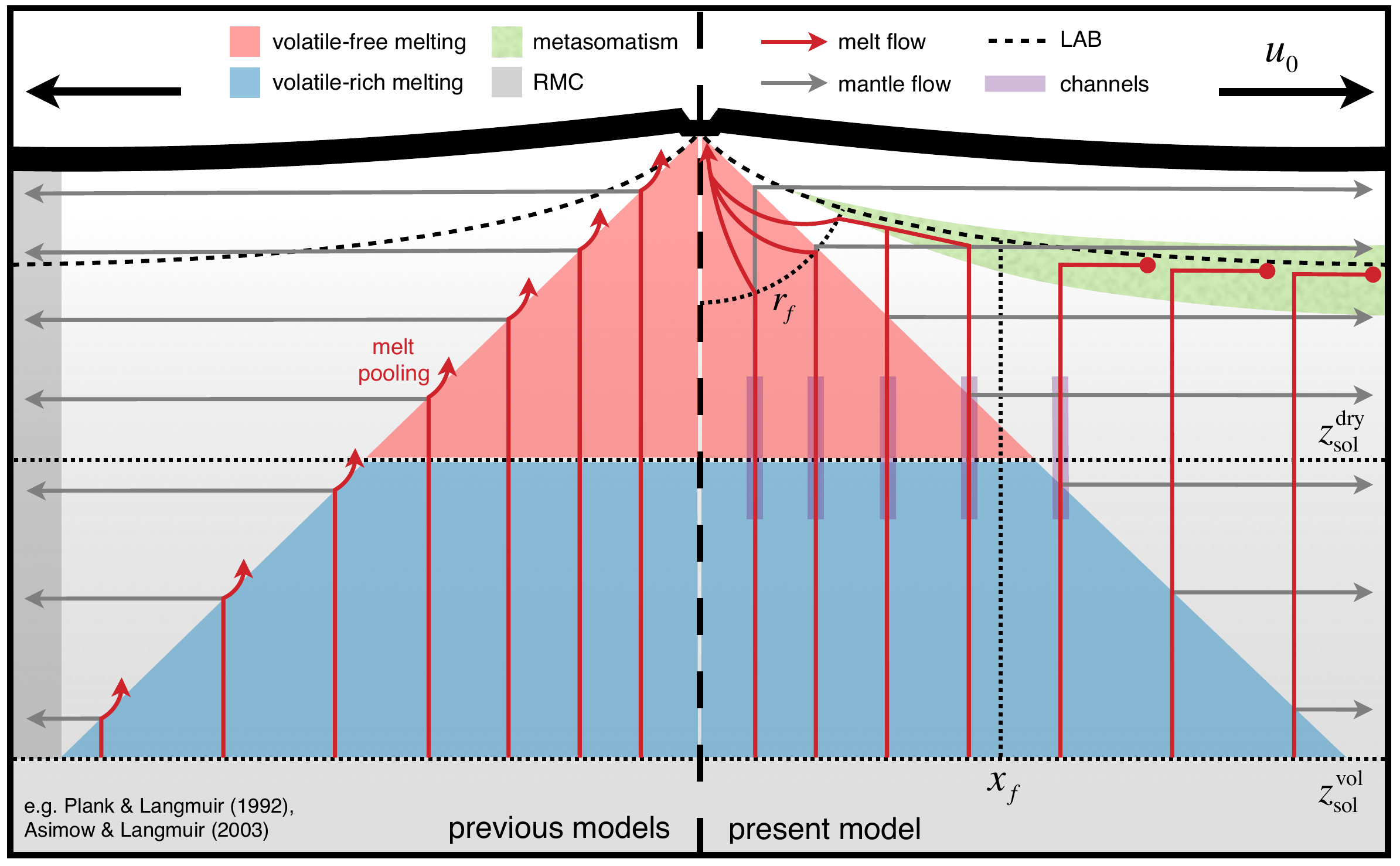}
  \caption{Summary of melt focusing beneath a MOR. Shaded areas show
    volatile-bearing (blue) and volatile-free (red) domains
    decompression melting; reactive channelling (purple); area of
    crystallisation and metasomatism (green). Dotted lines mark depths 
    of first melting for volatile-bearing, $z_\mathrm{sol}^
    \mathrm{vol}$, volatile-free mantle, $z_\mathrm{sol}^\mathrm{dry}
    $; lateral melt focusing distance $x_f$; active focusing radius 
    $r_f$. Melt (red) and solid (grey) flow paths are strongly 
    simplified. Left hand side presents simplified standard model, 
    compared to model presented in this study on the right hand side. 
    The key difference is the fate of deep, volatile-rich melt.}
  \label{fig:3}
\end{figure}

Volatile-rich melt is produced over a broad region beneath the
volatile-free melting regime. Extraction of these deep melts depends
on the efficiency of melt focusing towards the ridge axis. 
Interpretation of observed volatile contents of MORBs therefore
requires a model, whether implicit or explicit, of melt extraction
from the melting regime.  The canonical model of MORB petrogenesis
analyses melting and melt focusing in terms of the residual mantle
column~(RMC) --- the lateral solid outflow from the melting regime,
after steady-state melt extraction \citep[e.g.,][]{klein87,
  mckenzie88, plank92}.  Figure~\ref{fig:3} juxtaposes this model
(left-hand side) against a conceptual model derived from simulations
presented here (right-hand side).

\cite{plank92} and \cite{asimow03} calculate volume and composition of
extracted melt using the RMC (Fig.~\ref{fig:3}, left). All melt
produced within a chosen focusing distance from the ridge is assumed
to arrive at the axis, a concept referred to by the authors as ``melt
pooling.''  Volume and composition of extracted melt are calculated as
a weighted sum of melts produced at various pressures and degrees of
melting \citep{langmuir92}. Although most of the calculations in
\cite{plank92} assume perfect focusing of distal melts, they
recognised that ``it is especially difficult to envision efficient
extraction and focusing over hundreds of kilometers;'' they discuss
these difficulties in some detail. \cite{asimow03} do not consider the
limits of focusing; their calculations pool melts from the entire
melting regime by integrating over the full depth of the RMC.

In the present calculations, melt focusing is not an assumption but
rather a consequence of model dynamics. Careful analysis of
results suggests that focusing arises by two mechanisms, 
represented schematically on the right-hand side of Fig~\ref{fig:3} 
(c.f.~Suppl.~Figs~\ref{fig:S1}--\ref{fig:S3}).  Melt ascends by 
vertical porous flow driven by melt buoyancy.  As it approaches the 
thermal boundary layer along the LAB, crystallisation and 
increasing rock viscosity prohibit further vertical ascent. Instead, 
melt is diverted towards the axis, continuing as gravity-driven flow 
in a decompaction channel along the LAB \citep{sparks91, montesi11}. 
We refer to this first mechanism as passive focusing. The passive 
focusing distance $x_f$ arises from balancing rates of melt supply 
from beneath and loss to freezing into the base of the 
spreading plate; it is modulated by the slope of the LAB 
\citep[e.g.,][]{katz08, hebert10}.

Under the second mechanism, plate spreading induces a dynamic pressure
gradient that focuses melt transport toward the axis 
\citep{spiegelman87}; we refer to this as active focusing.  It extends 
radially from the ridge to a radius $r_f$.  Inside this radius, melt 
streamlines curve away from the thermal boundary on the ridge flanks, 
pointing down the dynamic pressure gradient towards the axis.  The 
focusing radius scales with the compaction length,
\begin{linenomath*}
  \begin{equation}
    \label{eq:1}
    \delta_c = \sqrt{\zeta K / \mu} \: ,
  \end{equation}
\end{linenomath*} 
where $K$ is the permeability, and $\eta$ and $\zeta$ the shear and 
compaction viscosities, assuming $\zeta\gg\eta$. In two-phase dynamics, 
pressure perturbations decay away from an applied forcing over a length 
scale of approximately the compaction length \citep{mckenzie84}. 
Depending on assumptions about permeability, compaction viscosity and 
melt viscosity, and for melt fractions of 0.1--1~wt\%, $\delta_c$ in the 
asthenosphere may be of order 1--100~km 
(see Appendix~\ref{sect:rheology}). 

Conductive heat flux to the surface causes cooling along the flanks of 
the melting regime. Some melt pathways will traverse these regions, 
where cooler temperatures cause deep fractional crystallisation.  As a 
consequence, volatiles (and other incompatibles) become more enriched 
in these melts.  Hence the volatile content of focused melts depends 
to some degree on their focusing pathways, which in turn are 
controlled by the two focusing mechanisms above.  Previous models did 
not include these effects or their consequences for volatile 
extraction.

For parameter ranges considered here, the active focusing radius is
typically smaller than the passive focusing distance. Thus $x_f$ marks 
the outer boundary of the melt focusing domain. The melt streamline in
Fig.~\ref{fig:1}(d) approximately marks the path of the most distal
melt still focused to the axis. We use the distance at which it 
intersects the base of the melting regime as a proxy for
$x_f$. \cite{katz08} proposed an independent measure of $x_f$ based on
integrated mass fluxes; it is termed the equilibrium focusing distance
$x_e$ (orange dotted, Fig.~\ref{fig:1}(d)). Both measures typically 
produce consistent estimates.  $x_f$ correlates well with
the lateral distance where the $u_0/3$ isopleth of solid upwelling 
rate intersects the volatile-free solidus depth (white line in 
Fig.~\ref{fig:1}(d)). At reference parameters, we find 
$x_f \approx 65$~km.

Model results suggest that the primary effects of volatiles --- deeper
onset of melting, a broader melting regime and reactive channelisation
--- do not have any significant effect on melt focusing.  Thus, only
volatile-rich melt produced in the central parts of the deep melting
regime reaches the MOR axis.  This result differs sharply from assumptions of 
\cite{asimow03} and similar models.  Furthermore, although volatiles cause 
channelisation of melt flow, this does not affect melt focusing: 
channels emerge in alignment with the prior direction of melt 
transport.  The main effect of channelised melt transport is instead
to increase the spatial and temporal variability of melt flux and 
composition.

Melt focusing that is independent of deep, volatile-rich melting
limits the efficiency of volatile extraction relative to previous
estimates \citep[e.g.,][]{asimow03}. Basalt extracted at the axis is
therefore dominated by low-pressure, high-degree melt. Most of the
low-degree melt produced in the distal wings of the volatile-rich
melting regime, conversely, migrates towards the LAB. There it is
collected into lenticular bodies with melt fraction of 5--50~wt\% at
20--60~km depth (Fig.~\ref{fig:1}(a)). These liquids metasomatise the
base of the lithosphere. As they crystallise and fractionate, their
volatile content increases while their freezing point decreases. We
discuss these effects in section~\ref{sect:lab}, below.

\section{Quantifying volatile extraction \label{sect:fluxes}}

\subsection{Melt production and focusing}
To quantify production and focusing of melt in MOR simulation
results we calculate the following rates of mass transfer per unit 
length of MOR axis: $\qb$, the flow rate of component mass into the 
base of the melting regime ($z_\mathrm{sol} = 160$~km); 
$\qm$, the rate of component mass transfer by melting over the full 
melting regime; $\qf$, the rate of component mass delivery to the 
ridge axis. Their calculation is discussed in 
Appendix~\ref{sect:app-a}. 

The ratios $\Rm$ and $\Rf$ are then introduced to quantify the mean
degree of melting beneath the ridge and the efficiency of melt
focusing to the axis.  The melt production ratio $\Rm$ is the
time-averaged melt production rate divided by the time-averaged mantle
inflow rate. The melt focusing ratio $\Rf$ is the time-averaged melt
focusing rate divided by the time-averaged melt production rate. These
ratios are calculated for each component and for the bulk mantle as
\begin{subequations}
  \label{eq:2}
  \begin{linenomath*}
    \begin{align}
      \label{eq:2a}
      \Rm &= \dfrac{ \langle \qm \rangle }{ \langle \qb \rangle }\:,\\
      \label{eq:2b}
      \Rf &= \dfrac{ \langle \qf \rangle }{ \langle \qm \rangle }\:,
    \end{align}
  \end{linenomath*}
\end{subequations}
where $\langle\cdot\rangle$ denotes a time average. 

Time-averaging is used to analyse the mean, long-term behaviour of the 
ridge apart from internal fluctuations. The time interval over which 
averages are taken corresponds to plate spreading distances between 
$x_0=180$~km and the full width of the model domain 
$x_1=W=300$~km. By that interval, plate spreading and solid upwelling 
have swept once through the melt regime, and visual inspection suggests 
that transients associated with model spin-up have disappeared 
(see Suppl.~Figs~\ref{fig:S4}--\ref{fig:S6}). Below, we discuss model 
output in terms of time-averaged rates $\langle q \rangle$ (and derived 
quantities) and their standard deviations $\sigma$ across this post-spin-
up time interval.

\begin{figure}[htb]
  \centering
  \includegraphics[width=\textwidth]{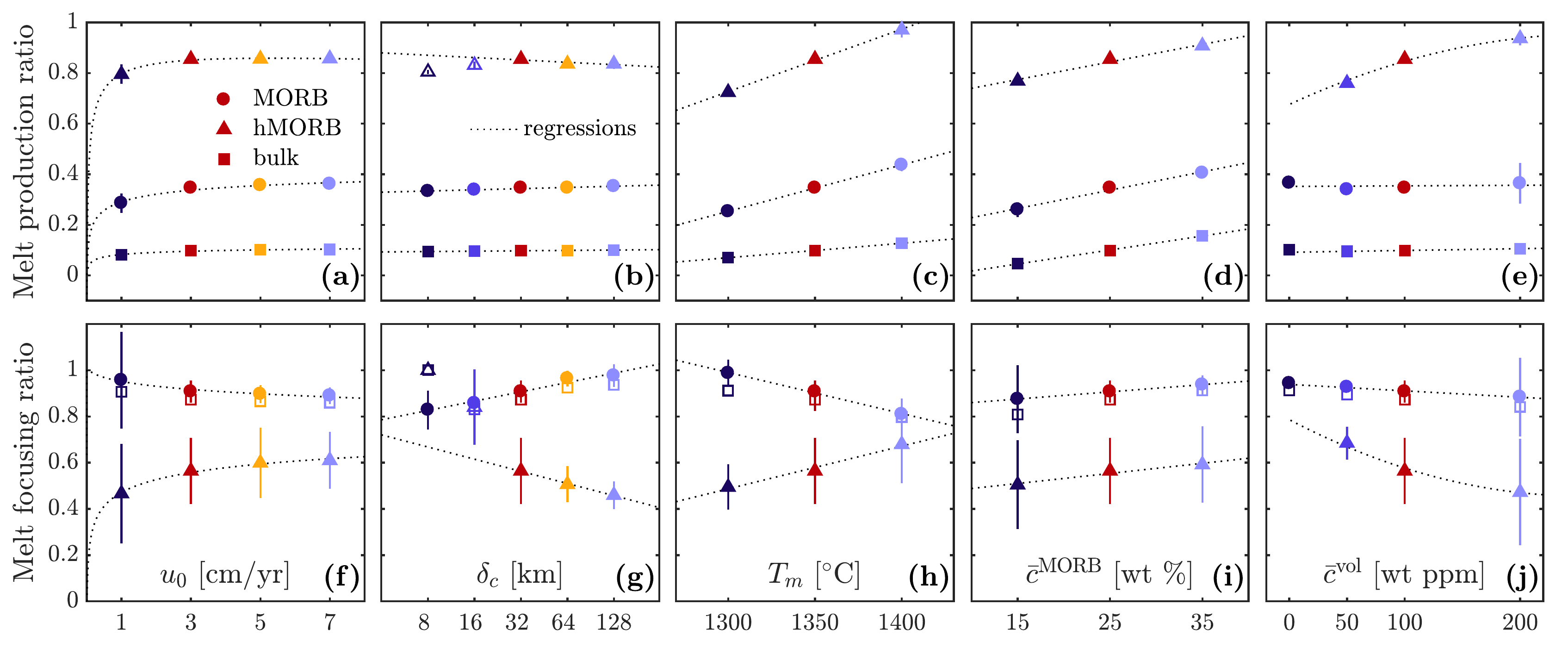}
  \caption{Magmatic ratios from \eqref{eq:2} as a function of
    parameter values. Time-averaged melt production ratios for MORB
    (circles), hMORB (triangles), and bulk mantle (squares). Results
    are plotted as a function of: \textbf{(a)} half-spreading rate;
    \textbf{(b)} compaction length; \textbf{(c)} potential
    temperature; \textbf{(d)} fertility; \textbf{(e)} volatile
    content. Time-averaged melt focusing ratios for MORB (circles),
    and hMORB (triangles), and bulk mantle (squares). Bars indicate 
    $\pm 2 \sigma$. Dotted lines are regressions of the 
    simulation output. Open symbols not included in regressions.}
  \label{fig:4}
\end{figure}

Figure~\ref{fig:4} shows time-averaged melt production and focusing 
ratios plotted against model parameters including spreading rate, 
compaction length, mantle temperature, fertility, and volatile 
content.  Parameter variations are applied one parameter at a time with 
other parameters held at reference values. Variations span a range of 
1--7~cm/yr in half-spreading rate, 8--128~km in compaction length (see 
Appendix~\ref{sect:rheology}), 1300--1400~\degC{} in mantle 
temperature, 15--35~wt\% in mantle fertility (MORB content), and 
0--200~ppm in volatile content (\water{} and \carbon{} added as hMORB 
and cMORB).

The bulk melt production ratio is $\sim$10\%. This is comparable to
the mean degree of melting discussed by \cite{asimow03}. $\Rm$
correlates positively with mantle temperature and fertility but is
largely independent of spreading rate (at least for slow to fast
spreading ridges), compaction length and volatile content. We do not
observe a lowering of $\Rm$ with the addition of volatiles, as
described for the mean degree of melting by \cite{asimow03}. This
difference comes from normalising $\Rm$ by mantle inflow at a fixed
depth (160~km) rather than the volatile-dependent depth of first
melting used by \cite{asimow03}.

The melt production ratios for the hMORB and MORB component illustrate
the difference between volatile-rich and volatile-free melt
production. While the former shows degrees of melting of 0.85--0.95,
the latter only reaches 0.25--0.4. The dissolution of the
incompatible, water-rich component from the partially molten
asthenosphere is therefore nearly complete in our models. Results for 
the carbonated component are similar to those for hMORB, and are 
therefore not shown in Fig.~\ref{fig:4}. The melts extracted at the 
ridge axis show no fractionation of water and carbon from their 1:1 
source ratio. This is simply due to the high mean degree of melting
that dominates the composition of extracted melts.

\cite{plank92} considered focusing of distal, volatile-rich melts to 
be physically implausible.  Here, we assess the efficiency of melt 
focusing and volatile extraction arising from dynamic simulations. 
Fig.~\ref{fig:4}(f)--(j) shows melt focusing ratios for the MORB and 
hMORB components.  80--100\% of MORB melt is delivered to the axis, 
whereas only 40--70\% of hydrated melt is focused. The remaining $\sim$60\% 
of dissolved water is transported towards the LAB and thus remains in 
the mantle. Volatile extraction efficiency is largely independent of 
spreading rate in panel~(f) and of mantle fertility in panel~(i). 
However, it decreases with increasing volatile content in panel~(j). 
The addition of volatiles enhances deep melting in the distal parts of 
the melting regime, yet the focusing distance is unaffected.  Thus, an 
increasing proportion of dissolved volatiles is not focused to the 
axis.

The striking anti-correlation between focusing of MORB and hMORB in
Fig.~\ref{fig:4}(g)--(h) shows the control of melt extraction pathways
on extracted melt compositions. With increasing compaction length,
active focusing becomes more important relative to passive
focusing. Melt streamlines curve away from the vertical toward the
ridge axis at greater depth (c.f.~Suppl.~Fig.~\ref{fig:S2}).  Melt is 
transported away from the thermal boundary layer and hence interaction 
with the cooler lithosphere is minimised. Active focusing thus leads 
to less volatile enrichment by fractional crystallisation as compared 
to passive focusing.  Less enrichment is also expected for other 
incompatible elements in the melt. Some of the compositional 
diversity of extracted melt may thus be attributed to different 
pathways of melt extraction, rather than to variations in source 
composition or degree of melting.

\subsection{Crustal thickness}
Predicted crustal thickness is an important model metric that can be
compared with observations (see Appendix \ref{sect:app-a}).
Figure~\ref{fig:5}(a)--(e) shows the time averaged crustal thickness
$\langle H_c \rangle$ plotted against the same parameters as in
Fig.~\ref{fig:4} (full time-series of $H_c$ in
Suppl.~Fig.~\ref{fig:S4}).  Crustal thickness as a function of
spreading rate shows the characteristic saturation towards fast
spreading ridges. For comparison, MOR crustal thickness data obtained
from seismics and rare-Earth-element inversions \citep{white01} are
plotted alongside simulation results. Motivated by the variation with
spreading rate, we fit the simulation output and observational data
with a log-cubic functions. The results are shown as
dotted and dashed lines, respectively, in panel~(a). Qualitatively, the simulation
results follow the same trend as the observational data. However, our
reference model result of $\sim$12~km crustal thickness is
considerably higher than the observed $\sim$7~km. We will return to
this point below.

\begin{figure}[th]
  \centering
  \includegraphics[width=\textwidth]{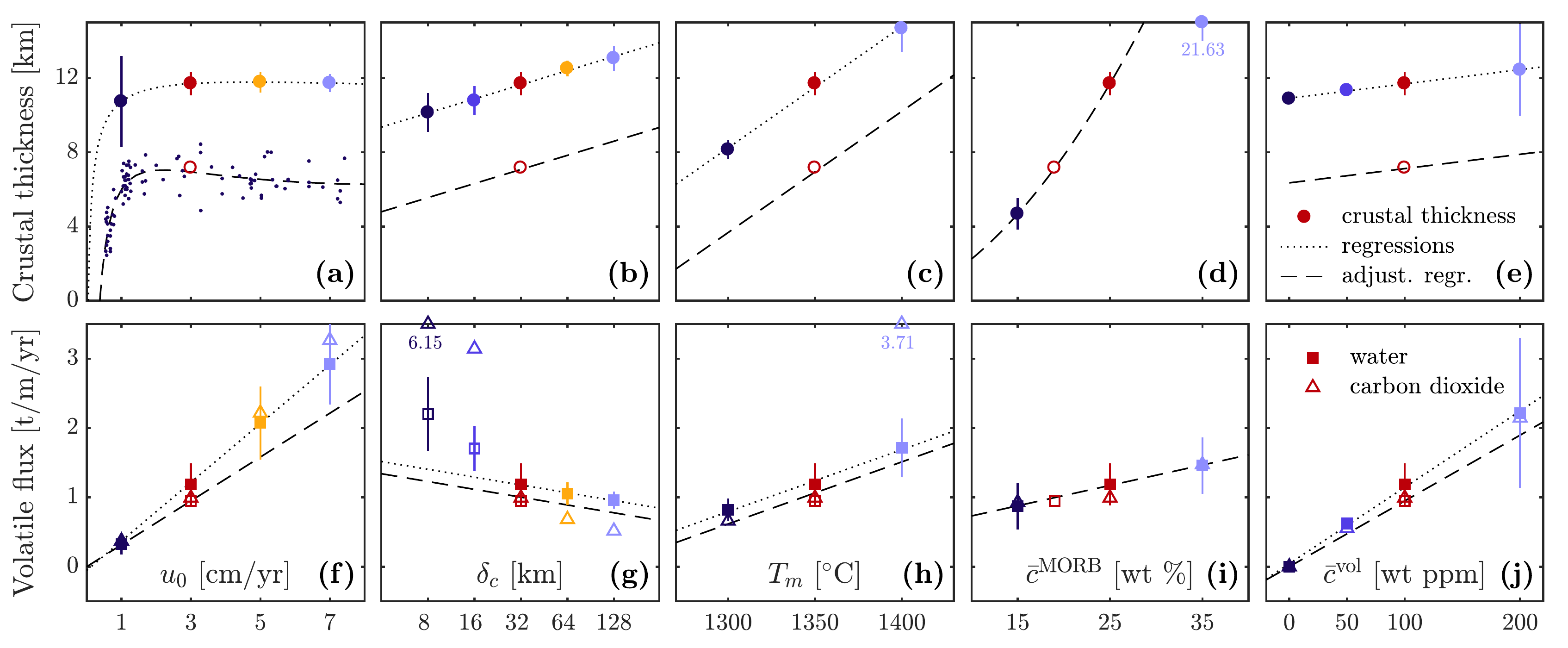}
  \caption{Time-averaged crustal thickness $\langle H_c \rangle$ and
    volatile extraction rate
    $\langle q^\mathrm{vol}_\mathrm{focus} \rangle$ as function of
    parameter values. Results for $\langle H_c \rangle$ (circles) are
    plotted as a function of: \textbf{(a)} half- spreading rate;
    \textbf{(b)} compaction length; \textbf{(c)} potential
    temperature; \textbf{(d)} fertility; \textbf{(e)} volatile
    content. Bars indicate $\pm 2 \sigma$. Dots in (a) are data from
    \cite{white01}. \textbf{(f)}--\textbf{(j)} show
    $\langle q^\mathrm{vol}_\mathrm{focus} \rangle$ for \water{}
    (squares), and \carbon{} (triangles) as function of same parameter
    variations.  Dotted lines are regressions of the simulation
    output; dashed lines are adjusted regressions to fit crustal 
    thickness data; open symbols not included in regressions.}
  \label{fig:5}
\end{figure}

Crustal thickness is shown to correlate with compaction length in
Figure~\ref{fig:5}(b). This is a consequence of more efficient melt
focussing with increasing compaction length. A log-linear regression
of the simulation output shows that a doubling of compaction length
leads to a $\sim$0.7~km increase in crustal thickness. Crustal
thickness also increases significantly with higher potential
temperature in panel~(c), and mantle fertility in panel~(d). A linear
regression of the results with temperature and a quadratic regression
with fertility show that 1~km of additional crust are produced by a
$\sim$16~\degC{} increase in $T_m$ or by a $\sim$1.5~wt\% increase in
$\bar{c}^ \mathrm{MORB}$, holding everything else constant.

Whereas these trends follow expectations from previous analyses, the
absolute melt production in our models is above observational
constraints. We use the regression from panel~(d) to calculate the
fertility reduction required to shift the fertility in the reference
model down such that the misfit with observational constraints is
reduced. The required shift in $\bar{c}^\mathrm{MORB}$ is
$-6$~wt\%. To check this shift, we include an additional simulation
with the MORB component concentration reduced to 19~wt\%. The result
is shown as open circles in panels~(a)--(e); as expected, the adjusted
crustal thickness is $\sim$7~km. Adjusted regression curves for other
parameters are shown as dashed lines.

The sensitivity of crustal thickness to mantle volatile content is
relatively small: 1~km of additional crust is produced by an increase
of $\sim$140~wt~ppm each of \water{} and \carbon{} (more than doubling
the reference value). In contrast, the temporal variability of crustal
thickness is highly sensitive to volatile content. It follows that
variability of magma supply at the ridge axis is not necessarily
diagnostic of source heterogeneity in mantle fertility or volatiles
--- all simulations are seeded with the identical mantle
heterogeneity. Rather, the temporal variability seen here is a result
of channelised melt transport and its associated intermittency in
space and time.

\subsection{Volatile extraction}
Rates of \water{} and \carbon{} extraction by focused magma at the 
axis are given by $q^\mathrm{H_2O}_\mathrm{focus}$, and $q^
\mathrm{CO_2}_\mathrm{focus}$ (see Appendix \ref{sect:app-a}). For 
\water{}, the extraction rate is interpreted as water supplied to the 
oceanic crust. Most of this water will remain contained within the 
crust but most of the \carbon{} focused to the ridge is degassed from 
the shallow magmatic system into the ocean--atmosphere reservoir.

Fig.~\ref{fig:5}(f)--(j) shows time-averaged volatile extraction rates
plotted against mantle parameters (full time-series of extraction
rates in Suppl.~Figs~\ref{fig:S5}--\ref{fig:S6}). Under most
conditions, \water{} and \carbon{} extraction rates are nearly
identical. As discussed above, volatile dilution by shallow,
high-degree melting precludes their fractionation. Furthermore we find
that the instances where the behaviour of \carbon{} and \water{}
differ are best explained by errors associated with insufficient
numerical resolution. Simulations with $\delta_c<32$~km have
significant mass conservation errors for the incompatible
components. Hence cMORB is more strongly affected while hMORB remains
mostly well resolved. MORB and dunite components are not affected (see
Suppl.~Figs~\ref{fig:S5}(f)~\&~\ref{fig:S6}(f) for resolution test results). 
We therefore proceed with discussion of the hMORB component and assert 
that these results are a meaningful proxy for \carbon{} extraction.

Volatile extraction rates increase linearly with half-spreading rate 
in panel~(f). At reference parameters the extraction rate is $\sim
$1200~kg/m/yr. In the model with shifted reference fertility it is 
reduced to $\sim$950~kg/m/yr (open square). A linear regression 
through the origin and this shifted reference rate (dashed line) 
has an increase in volatile extraction rate of $\sim $320~kg/m/yr 
for every 1~cm/yr increase in $u_0$. This sensitivity is comparable to 
idealised models of MOR carbon degassing by \cite{burley15}. The 
correlation with spreading rate is the result of increased melt 
production, while volatile concentration in the magma remains 
relatively stable (c.f. time-averaged melt compositions in 
Suppl.~Fig.~\ref{fig:S7}). 

The decrease of extraction rates with increasing compaction length in 
panel (g) is, again, associated with the shift toward more active 
focussing. As melt is diverted away from the sub-lithospheric 
crystallisation front, deep fractional crystallisation and the 
consequent enrichment of incompatibles is reduced. Results at small 
$\delta_c$ suggest insufficient numerical resolution. A log-linear 
regression of the well-resolved results at higher $\delta_c$ has a 
drop in volatile extraction rate of $\sim$110~kg/m/yr for a doubling 
of $\delta_c$. An increase in mantle temperature in panel (h) and 
fertility in panel~(i) both lead to a moderate increase in volatile 
extraction, as the deepening and widening of the volatile-free 
melting regime allows some more distal volatile-rich melt to be focused. 
Linear regressions of these results show that a rate 
increase of 100~kg/m/yr is produced by either a temperature increase 
of $\sim$14~\degC{} or an additional $\sim$3.4~wt\% mantle fertility. 

Volatile extraction at the ridge axis is proportional to the mantle 
volatile content in panel~(j). Regression through the origin and the data 
point from the adjusted fertility reference simulation shows that an 
additional 10~wt~ppm volatiles in the mantle gives an increase of $\sim
$95~kg/m/yr in ridge extraction rate. The signature of increased 
channelised transport at higher volatile contents is expressed by higher 
variability of extraction rates. The relative amplitude of these variations is 
up to 50\%.  Again, this variability is not caused by source heterogeneity 
or variable degrees of melting. Rather it is the signature of reactive melt 
transport in the asthenosphere.

\section{Simulation-based estimates of global MOR volatile output 
	\label{sect:rates}}

\subsection{Distribution along the global MOR system}
The simulations presented here can provide a novel perspective on the
globally integrated MOR output.  Using the regressions of simulation
output in Fig.~\ref{fig:5} we interpolate crustal thickness and
volatile extraction rate to conditions found along the global MOR
system. However, in reading the results, one should bear in mind the
limitations of this method: each regression curve was obtained for an
isolated parameter variation in a complex, non-linear system; the
natural system may be characterised by significant covariation between
parameters. Despite this limitation, we expect to gain
insights into possible global patterns as well as globally integrated
magnitudes of MOR output.

\begin{figure}[htb]
  \centering
  \includegraphics[width=\textwidth]{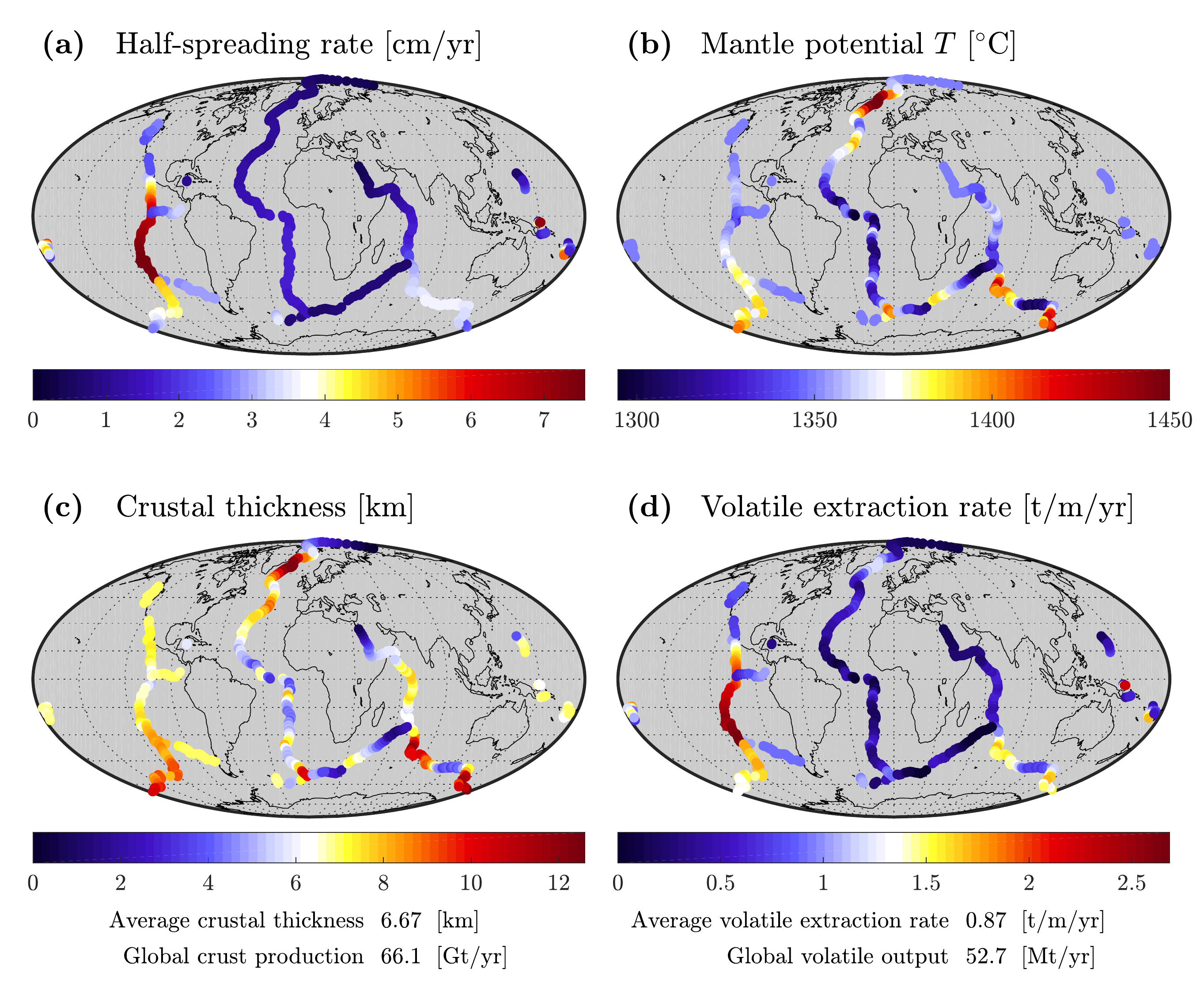}
  \caption{MOR magmatic flux estimates for global ridge system with 
    $\bar{c}^\mathrm{H_2O}= 100$~ppm:
    \textbf{(a)} Half-spreading rates taken from \cite{gale14};
    \textbf{(b)} Mantle potential temperature extrapolated from
    \cite{dalton14}; \textbf{(c)} Crustal thickness, and \textbf{(d)}
    volatile extraction rate calculated from regressions of simulation 
    outputs. Each data point represents one ridge segment in
    \cite{gale14} catalogue.}
  \label{fig:6}
\end{figure}

Conditions along the global MOR system are available from a catalogue 
of ridge segments composed by \cite{gale14}, which gives the location, 
length, and spreading rate of each segment. \cite{dalton14} obtained 
estimates of potential temperature beneath segments on the basis of 
inversion of observed ridge depth, S-wave speed, and MORB 
geochemistry. We use those estimates to assign mantle temperature for 
each ridge segment in the \cite{gale14} catalogue. We shift the 
temperatures by -25~\degC{} to align their average temperature to our 
reference temperature of 1350~\degC. For ridge segments not included 
in \cite{dalton14}, we interpolate between the nearest available 
constraint and 1350~\degC.  Figure~\ref{fig:6}(a)--(b) shows half-
spreading rate and mantle potential temperature plotted on the 
mid-point coordinates of each ridge segment in the combined catalogue.

Global average values for mantle fertility, volatile content and
compaction length are approximately known. Regions of increased volatile 
content are typically linked to nearby hotspots (e.g., Azores 
\citep{asimow04}). Yet, emerging information about these variations along 
the ridge system is still sparse \citep{levoyer17}. Therefore, we will  
investigate the range of $100 \leq \bar{c}^\mathrm{H_2O} \leq 200$~ppm, along with reference values for $\delta_c=32$~km, $\bar{c}^\mathrm{MORB}=19$~wt\% (the adjusted value). We can now use the 
adjusted regressions in Fig.~\ref{fig:5} to compute crustal 
thickness and volatile extraction rate for each segment in the catalogue.

The predicted global distribution for $\bar{c}^\mathrm{H_2O}= 100$~ppm, is shown in Fig.~\ref{fig:6}(c)--(d). The pattern of crustal 
thickness in panel~(c) mostly follows the pattern of mantle temperature 
from \cite{dalton14}. The thickest crust is found in regions influenced by
hot-spots, most pronounced beneath Iceland. There crustal thickness
exceeds 12~km. The global average crustal thickness is 6.67~km. The
lowest values are found at the eastern end of the Southwest Indian
Ridge, at $<2$~km. The distribution of volatile extraction rate, in
contrast, shows a strong dependence on spreading rate, with only a
weak influence of mantle temperature. The highest volatile extraction
rates of $>2.5$~t/m/yr occur along the East Pacific Rise. The global
average rate is 0.87~t/m/yr. For cool, slow-spreading ridge sections 
such as parts of the Mid-Atlantic and Southwest Indian Ridges, values 
are $<0.25$~t/m/yr.

\begin{figure}[ppp]
  \centering
  \includegraphics[width=\textwidth]{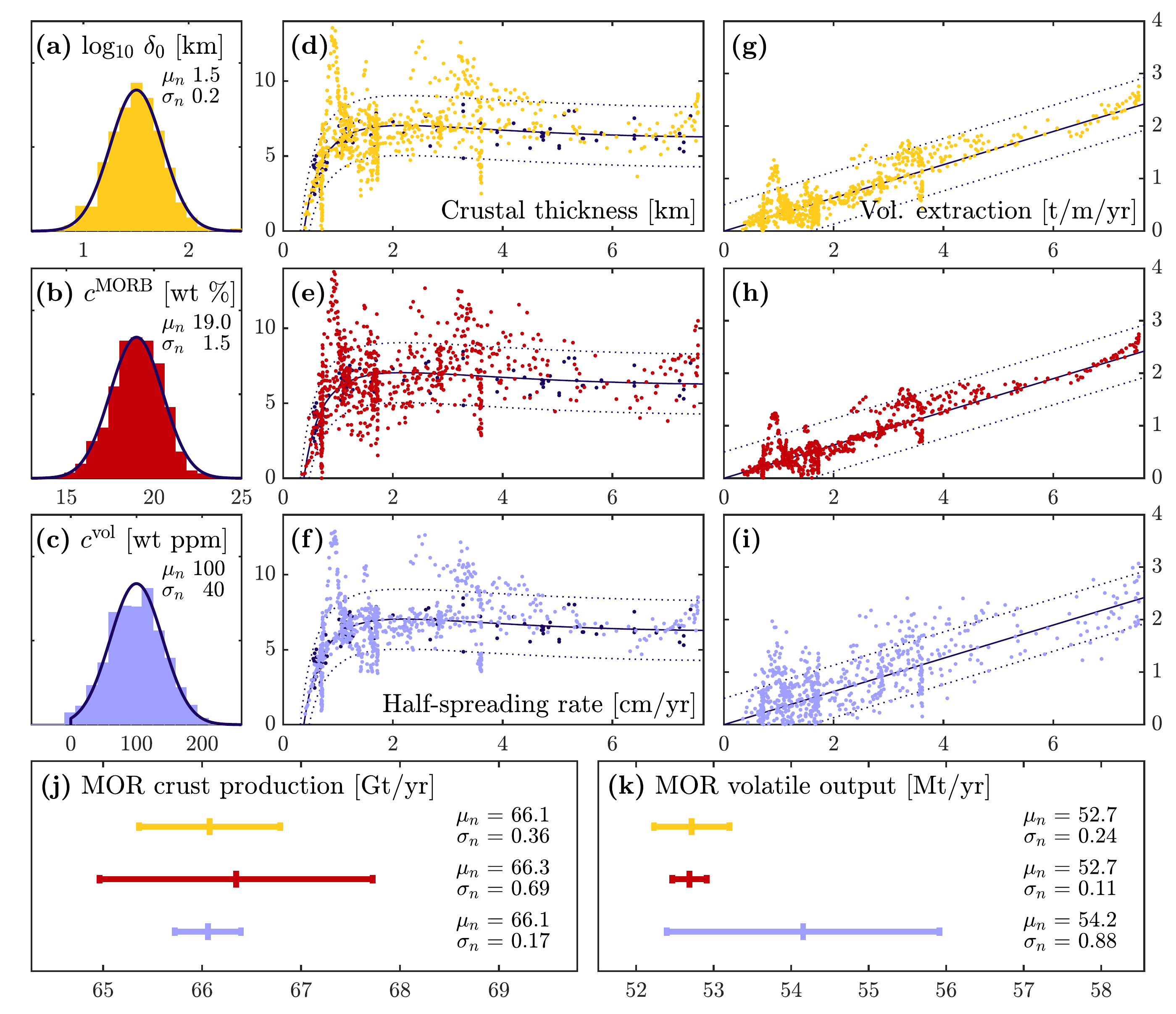}
  \caption{Globally integrated MOR magmatic flux estimates. Parameter
    distributions (line) and example histogram (bars) for segment-wise
    sampling of \textbf{(a)} compaction length, \textbf{(b)} mantle
    fertility, and \textbf{(c)}, volatile content. Crustal thickness
    \textbf{(d)--(f)}, and volatile extraction rate \textbf{(g)-(h)}
    calculated from fitting functions for example parameter
    distribution. \cite{white01} data points for comparison (black);
    Lines show adjusted fitting functions for crustal thickness with
    best fit parameters (solid), and $\pm$ 2 km variation (dotted),
    and for extraction rate with best fit parameters (solid) and
    $\pm$ 0.5 t/m~yr variation (dotted). Normal distributions fitted
    to population of globally integrated output from 1383
    sampling instances (j)--(k). Colours denote individually sampled
    problem parameters.}
  \label{fig:7}
\end{figure}

\subsection{Range of global MOR output}
Globally integrated MOR output can be obtained from the distributions
above. For crust production, we multiply crustal thickness by 
$2 u_0 \rho_c$ and integrate over the length of ridge segments. The 
resulting projection of global crust production is 66.1~Gt/yr---
equivalent to 22~km\textsup{3}/yr. Similarly, integrating the volatile 
extraction rate over the length of ridge segments after 
\cite{burley15} predicts the global MOR volatile output to be 
52.7~Mt/yr of \water{} and \carbon{} each for reference volatile 
content. Repeating the global calculation above for a mean mantle volatile
content of 200 ppm gives a global volatile output of 110.2~Mt/yr. The 
corresponding global crust production is 73.1~Gt/yr 
(24~km\textsup{3}/yr). The average volatile concentration 
in (undegassed, undifferentiated) focused melt is calculated by dividing the 
global volatile extraction by the crust production rate. The calculated 
values are 0.08 and 0.15~wt\% for 100 and 200~ppm \water{} and 
\carbon{} in the mantle. According to our calculations, measured \water{} 
(0.13--0.77~wt\%) and reconstructed \carbon{} (0.066--5.75~wt\%) 
concentrations in MORB by \cite{cartigny08} would map back to global 
mean concentrations of $>$170~ppm \water{} and $>$80~ppm 
\carbon{}; the upper limits would map to values well above 
1000 ppm.

In addition to these leading-order global figures, we estimate the
uncertainty of global MOR output arising from uncertainty in
geographic, segment-wise variations of parameters that are not listed
in the above catalogue.  To this end we assume that mantle fertility,
volatile content and compaction length are not constant in the upper
mantle, but are distributed over a plausible range. Black curves in
Figure~\ref{fig:7} (a)--(c) show distributions of compaction length,
fertility, and volatile content that are proposed to represent
variation in the asthenosphere beneath the ridge system. Each
distribution is centred around the reference values above. We assume
compaction length has a log-normal distribution across a range of
$\sim$10--100~km ($ \mu_n \pm 2 \sigma_n$); fertility is normally
distributed with a range of $\sim$16--22~wt\%; volatile content is
normally distributed with a range of $\sim$20--180~wt~ppm 
(truncated at zero).

A population of instances of global MOR output is produced by
repeating the following three steps. First, one of the three
parameters is randomly sampled according to its distribution to obtain
a value for each ridge segment. A histogram of a sample is shown for
each parameter in Fig.~\ref{fig:7}(a)--(c). Second, segment-wise
crustal thickness and volatile extraction rate are computed from
regressions as above.  Calculated values for sampled parameter sets in
panels (a)--(c) are plotted against spreading rate in panels
(d)--(i). Third, instances of globally integrated MOR output are
calculated by integrating over the length of ridge segments. These
three steps are repeated separately for each of these three 
parameters until each parameter-specific population of instances is 
fitted by a normal distribution to a relative tolerance of $10^{-3}$.

The populations for crust production and volatile output 
obtained from $\gtrsim10^3$ instances of statistically sampled 
parameter sets are shown in Fig.~\ref{fig:7}(j)--(k). The ranges of 
outcomes are shown as $\mu_n \pm 2 \sigma_n$. Global crustal 
production is most sensitive to mantle fertility variations with a 
range of 65--68~Gt/yr.  Global volatile extraction rate is most 
sensitive to mantle volatile content with a range of 52--56~Mt/yr. 
Variations in compaction length and mantle fertility have second-order 
effects on the range of volatile output.

These projected values of global volatile export from the mantle fall
within the range of previous estimates, particularly in comparison to
previous estimates of \carbon{} emissions from MOR degassing
\citep{resing04, cartigny08, dasgupta10, kelemen15}. However, 
limitations of the present method should moderate our confidence 
in the produced estimates of MOR volatile output. The validity of the 
physical model and its numerical implementation is limited. For
example, a more realistic rheology depending on stress and dynamically
evolving grain size could have important effects on melt segregation
\citep{turner15}; the model of mantle melting is based on
simplified thermodynamics that are, at best, a crude approximation of
full mantle petrogenesis; and the simulations skirt the limits of
spatial resolution, introducing potentially disruptive mass
conservations errors for incompatible species.  Furthermore, using
single-parameter regressions of simulation output neglects possible
non-linear effects of parameter co-variations; polynomial degrees
for regressions were chosen for convenience, not consistency with
physical processes.

Nevertheless, the present method produces estimates of MOR volatile
output that are consistent with dynamic simulations taking into 
account conservation of mass, momentum and energy. The models are 
calibrated to reproduce accepted features of MORB petrogenesis to 
leading order. Model parameters are tuned with the most recent 
observational constraints on the global MOR system, where available. 
The present estimates of global MOR volatile output are consistent 
with available constraints on extraction processes. Our method 
estimates MOR output for a range of global mean mantle properties and 
it quantifies the uncertainty associated with plausible 
regional distributions around reference conditions.

\section{Deep volatiles and the LAB \label{sect:lab}}
From the above it is evident that not all melt produced beneath a MOR
is focused to the ridge axis. Instead, a significant fraction of the
deep, volatile-rich melt is transported towards the oceanic LAB. As
seen in Fig.~\ref{fig:1}, this transport creates remnants of reactive
channels bent away from the axis through plate motion. Over time such
channels become stagnant sub-horizontal melt bodies stacked along the
LAB. This feature of our results could help explain the sharp negative 
shear-wave velocity contrasts identified along the oceanic LAB in 
recent studies \citep{kawakatsu09, schmerr12, stern15}. We find that
such melt lenses, which contain melt fractions of 5--25\% (some up to 
50\%), are typically located along the 1200~\degC{} isotherm. Some of this 
stalling melt could be further extracted by dike propagation 
\citep[e.g.][]{havlin13}, an effect not currently included in our models. 
With ongoing crystallisation, volatiles are enriched in the fractionated liquid; 
the oceanic LAB is metasomatised by these volatile-rich liquids. In contrast 
to simpler models \citep[e.g.][]{plank92, hirth96, asimow03}, the volatile 
content with depth beneath the ocean floor (i.e., the RMC) does not 
reflect either a fractional or batch melting residue. Instead, the depth 
profile of volatiles is a result of complex transport processes. In particular, 
channelised melt transport leads to a heterogeneous distribution of 
volatiles along the LAB as it moves away from the ridge axis.

\begin{figure}[htb]
  \centering
  \includegraphics[width=\textwidth]{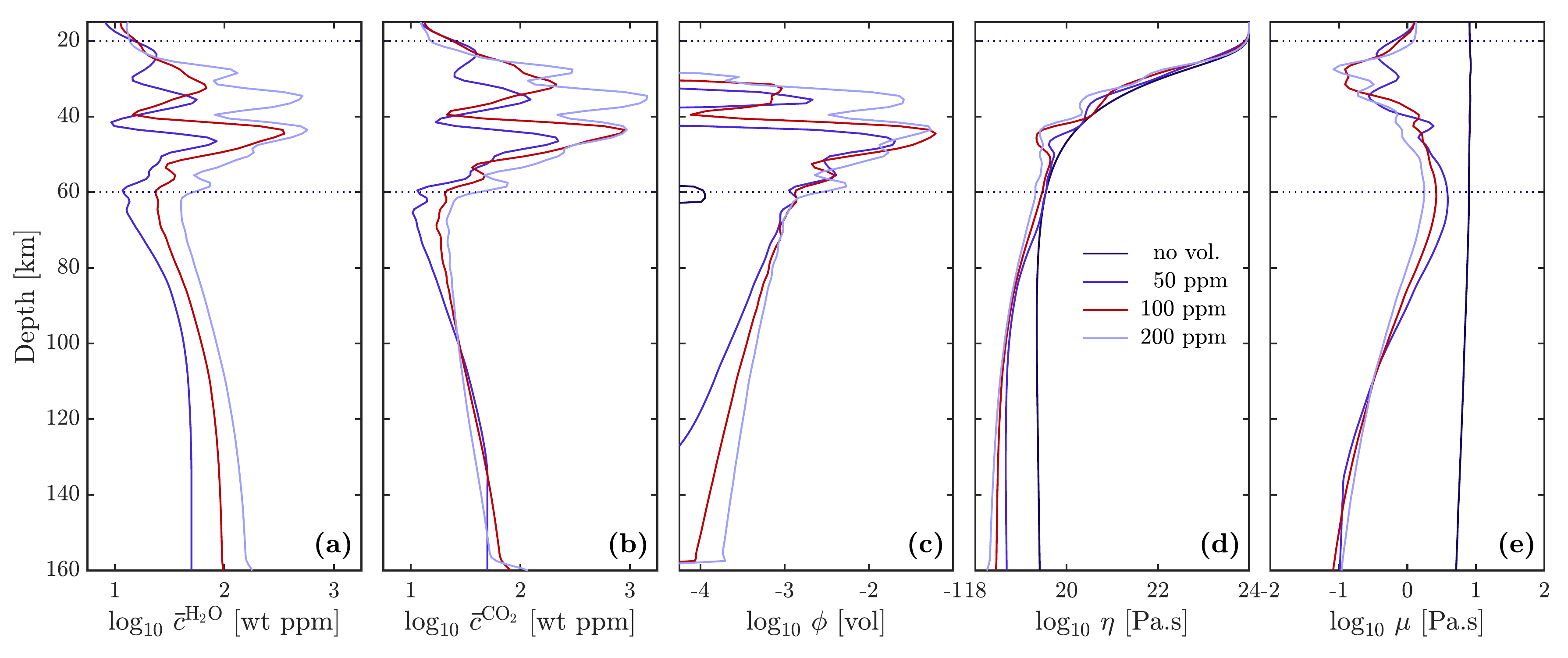}
  \caption{Averaged depth profiles of volatile and melt content, and
    rheology along far flank of ridge. Depth profiles obtained by
    horizontal averaging over interval from 200--300 km distance from
    axis, and triangular moving average along profile. Water and
    carbon dioxide bulk concentrations \textbf{(a)} \& \textbf{(b)};
    melt fraction \textbf{(c)}; rock viscosity \textbf{(d)}; melt
    viscosity \textbf{(e)}. Given for simulations with 0 (black), 50
    (blue), 100 (red), 200 (light blue) wt ppm initial volatile
    content. Where no melt present, melt viscosity is found from
    fractional melt composition.}
  \label{fig:8}
\end{figure}

Figure~\ref{fig:8} shows depth profiles produced by horizontally
averaging simulation results in the interval from 180 to 300~km from
the axis (6--10 Myr plate age).  These profiles are then filtered with
a vertical moving-average to emphasise structures larger than
$\gtrsim$5~km. Figure~\ref{fig:8} shows results from simulations with
various volatile contents.  Moving upwards from 160 to 60~km depth,
both \water{} and \carbon{} follow a trend of depletion.  Above is 
found an enriched region from 60 to 20~km depth. The enrichment is
layered and reaches concentrations up to 1000~ppm; it is stronger for
\carbon{} than for \water{}, which indicates fractional
crystallisation of accumulated low-degree melts.  This layered band of
metasomatised mantle coincides with the top boundary of the partially 
molten domain.  Below 60 km depth, melt is stable at low fractions of
0.01--0.1\%. Along the thermal LAB at $\sim$60--50~km depth, melt is
accumulated into decompaction layers with horizontal averages upwards
of 1\% melt. Some remains of crystallising melt are found to a depth
of 30 km. It is conceivable that such melt layers are a source of
off-axis seamount volcanism.

Volatile enrichment along the LAB has consequences for rock
viscosity, as shown in Fig.~\ref{fig:8}(d). Viscosity is reduced
by water; a gradual increase in viscosity above the base of the
melting regime reflects volatile depletion by melt extraction. Along
the band of metasomatism, however, rehydration of the solid creates
layers with up to an order of magnitude reduction of viscosity. These
represent a structural heterogeneity at the base of the lithosphere. 
Such rheological structure could help explain the sharp decrease in shear-
wave speed observed along the oceanic LAB \citep[see][and references 
therein]{olugboji16}. Our results support the current hypothesis that these 
signals reflect the presence of partial melt and/or hydrated material along 
the LAB. Furthermore, a metasomatised oceanic lithosphere has recently 
been evoked as the potential source of enriched alkaline ocean island 
basalts \citep{workman04, pilet11} and highly alkaline petit-spot 
volcanism \citep{hirano11}.

Lastly, the depth profiles of magma viscosity in Fig.~\ref{fig:8}(e)
show a clear signal of lubrication by volatile enrichment. The signal 
is seen in deep primitive as well as in shallow fractionated melt. 
This viscosity variation promotes melt segregation from the mantle 
residue even at small melt fractions $<0.1$~wt\%. An appropriate 
liquid viscosity model with volatile content is thus crucial for 
understanding the transport of volatiles in the mantle beneath 
mid-ocean ridges.

%% CONCLUSIONS The main conclusions of the study may be presented in
%% a short Conclusions section, which may stand alone or form a
%% subsection of a Discussion or Results and Discussion section.

\section{Summary and conclusions  \label{sect:conclusions}}
We present dynamic models of thermochemically coupled magma/mantle
dynamics with volatiles beneath a mid-ocean ridge. The presence of
volatiles in low concentrations in the mantle source causes deep,
low-degree, volatile-rich melting. Due to the corrosive effect of
volatile-rich melt on silicates with decreasing pressure, the flux of
deep melt into the volatile-free melting regime triggers reactive
channelisation. We find that melt focussing to the axis is not
extended by channelised flow. Rather, channelling introduces
spatial and temporal heterogeneity in flux and melt composition at the
ridge axis.

We investigate volatile extraction from the asthenosphere across a
range of parameters including spreading rate, compaction length,
mantle temperature, fertility and volatile content. Volatile
extraction depends on the rate of melt delivery to the axis, as well
as the concentration of volatiles therein. The efficiency of melt
focusing increases with compaction length, mantle temperature and
fertility. The concentration of volatiles in that melt depends on the
melt extraction pathways and, in particular, whether they approach the
thermal boundary layer along the LAB, where fractional crystallisation
leads to volatile enrichment.
 
We calculate crustal thickness and volatile extraction rates for
individual ridge segments in a global catalogue using regressions of
simulation outputs and crustal thickness data. The resulting 
distribution of projected crustal thickness is mainly controlled by 
mantle temperature, whereas the distribution of volatile extraction 
rate dominantly reflects spreading rate. Global integration gives 
values of 66--73~Gt/yr (22--24~km\textsup{3}/yr) for global crust 
production and a global volatile output of 53--110~Mt/yr, 
corresponding to mantle volatile contents of 100--200~ppm. The 
uncertainty of these estimates relating to the uncertain distribution 
of mantle conditions along the global ridge system is assessed using 
statistical sampling from plausible distributions of compaction 
length, mantle fertility and volatile content. The resulting range of 
uncertainty for parameters distributed around reference mantle 
conditions is 65--68~Gt/yr for crust production and 50--55~Mt/yr for 
volatile output.

Finally, and in contrast to \cite{asimow03}, our models suggest that 
up to half of dissolved volatiles that cross into the melting regime 
are not focused to the axis, but are instead extracted towards the 
flanks of the ridge, where they metasomatise the mantle along the LAB. 
The resulting rheological heterogeneity caused by water-weakening of
mantle rock could explain seismic signals interpreted as evidence of
the oceanic LAB depth. Our results show bands of metasomatised rock at 
a depth range of 20--60~km at spreading ages of 6--10~Myr, consistent
with LAB-related signals in seismic studies.

\subsection*{Acknowledgements}
The research leading to these results has received funding from the
European Research Council under the European Union's Seventh Framework
Programme (FP7/2007--2013)/ERC grant agreement number 279925. Hirschmann acknowledges support from NSF grant EAR1426772. The
authors thank the Isaac Newton Institute for Mathematical Sciences for
its hospitality during the programme Melt in the Mantle which was
supported by EPSRC Grant Number EP/K032208/1. Katz is grateful for the
support of the Leverhulme Trust. The authors further thank the
Geophysical Fluid Dynamics group for access to the BRUTUS cluster at
ETH Zurich, Switzerland. Thanks to R.~White, C.~Dalton, and A.~Gale
for making their data available.

%% BIBLIOGRAPHY
\section*{References}
\bibliographystyle{elsarticle-harv} \bibliography{manuscript}

\cleardoublepage

\appendix
\renewcommand{\thesection}{\Alph{section}} \setcounter{section}{0}
\renewcommand{\thetable}{A\arabic{table}} \setcounter{table}{0}
\renewcommand{\thefigure}{A\arabic{figure}} \setcounter{figure}{0}

\section{R\_DMC: Equilibrium and reaction model \label{sect:thermodyn}}
The following is a brief summary of the R\_DMC method, which provides
a model of thermodynamic equilibrium and linear kinetic reactions for
multi-component compositional models of mantle petrogenesis. A
detailed description of the method is found in \cite{keller16}. The
method provides closure conditions for reaction rates $\Gi$ needed to
couple conservation equations for energy, phase and component mass in
a multi-phase, multi- component reactive flow model. It consists of
the following four steps:

(\textit{i}) Determine partition coefficients $\Ki$ for each effective 
component ($i=1,...,n$) at given $P,T$-conditions. The partition 
coefficients are functions of $T$ and $P$, following a simplified form of 
ideal solution theory \citep{rudge11}. Effective components are chosen to 
approximate the petrological processes of interest within a framework of 
tractable complexity. Here, we represent mantle melting 
with four effective components ($n=4$): dunite and mid-ocean ridge 
basalt as the residual and product of volatile-free mantle melting, and 
hydrated and carbonated basalt as the products of hydrated and 
carbonated silicate melting at depth.

(\textit{ii}) For a given bulk composition $\cbar = \phi \cli + \phis \csi$, 
determine the unique melt fraction $\phieq$ for which all components are 
in equilibrium according to $K^i$. We do so by combining the lever rules 
for each component, partitioned between the phases according to $K^i$, 
with the constraint that all component concentrations in each phase must 
sum to unity. Thus we formulate the implicit statement
\begin{linenomath*}\begin{align}
		\label{eq:EquilibriumMelt}
		\sum\limits_{i=1}^{n} \dfrac{\cbar}{\phieq / \Ki + (1-\phieq)} - \sum\limits_{i=1}^{n} \dfrac{\cbar}{\phieq + (1-\phieq) \Ki} = 0  \: ,
\end{align}\end{linenomath*}
which we solve numerically for $\phieq$.

(\textit{iii}) Find the equilibrium phase compositions $\cseq$ (solid), $\cleq$ (liquid) at equilibrium melt fraction $\phieq$ by lever rule	
\begin{linenomath*}\begin{align}
		\label{eq:EquilibriumComp}
		\cseq &= \dfrac{\cbar}{\phieq / \Ki + (1-\phieq)}, \\\nonumber
		\cleq &= \dfrac{\cbar}{\phieq + (1-\phieq) \Ki} \: .
\end{align}\end{linenomath*}

(\textit{iv}) Use linear kinetic constitutive laws to determine equilibration reaction rates $\Gi$ driven by compositional disequilibria
\begin{linenomath*}\begin{align}
	\label{eq:CompReactDeriv}
	\Gi = \RG (\phieq \cleq - \phi \cli) \: .
\end{align}\end{linenomath*}
The linear kinetic rate factor $\RG = \rho_0 / \tau_{\G}$ ensures that equilibration is applied over a specified reaction time scale $\tau_{\G}$, with $\rho_0$ the reference solid density. The net melting rate $\G$ is the sum of all component rates $\Gi$. These reaction rates drive the system toward chemical equilibrium.

\section{Rheology with volatiles \label{sect:rheology}}
The viscosity of mantle rock is taken as diffusion creep of olivine
with melt- and water-weakening \citep{hirth03, mei02}. Constitutive
laws for shear and compaction viscosities $\eta,~\zeta$ are written
in standard form as a function of lithostatic pressure $P$, 
temperature $T$, grain size $a_0$ (assumed constant and uniform), 
water content $C_H$, and melt fraction $\phi$:
\begin{subequations}
  \label{eq:A1}
  \begin{linenomath*}
    \begin{align}
      \label{eq:A1a}
      \eta^\mathrm{dry} &= A_0^\mathrm{dry} a_0^3 \exp 
        \left( \dfrac{E_a + P V_a}{RT} - \alpha \phi \right) \: , \\
      \label{eq:A1b}
      \eta^\mathrm{wet} &= A_0^\mathrm{wet} C_H^{-1} a_0^3 \exp 
        \left( \dfrac{E_a + P V_a}{RT} - \alpha \phi \right) \: , \\
      \label{eq:A1c}
      \eta &= \min(\eta^\mathrm{dry},\eta^\mathrm{wet}) \: , \\
      \label{eq:A1d}
      \zeta &= r_\zeta \: \eta \: \phi^{-1} .
    \end{align}
  \end{linenomath*}
\end{subequations}
The parameter $r_\zeta$ represents the poorly constrained ratio
of compaction to shear viscosity.  $C_H$ is obtained by converting
dynamically calculated water concentrations in the solid phase to the
appropriate units (1~ppm \water{} equals 5.4~H/($10^6$ Si), 
c.f.~\cite{hirth96}).

The viscosity of the volatile-bearing silicate melt is taken to depend
on composition: weakening with increasing concentration of \water{} 
and \carbon{}, stiffening with increasing dissolved silica. The 
consitutive law is log-linear:
\begin{linenomath*}
  \begin{align}
    \label{eq:A2}
    \mu = \mu_0 \prod\limits_{i=1}^n \lambda^i \cli \: .
  \end{align}
\end{linenomath*}
Factors $\lambda^i$ are chosen such that melt viscosity varies between
0.01~Pa-s for a volatile-rich melt to 10~Pa-s for a volatile-poor
melt. Permeability depends on grain size and melt fraction according
to the Kozeny-Carman relation, 
\begin{linenomath*}
  \begin{align}
    \label{eq:A3}
    K = \dfrac{a_0^2}{100} \dfrac{\phi^3}{(1-\phi)^2}.
  \end{align}
\end{linenomath*}

With the above constitutive laws the compaction length \eqref{eq:1} 
depends on temperature, melt fraction and composition, which are 
features of the solution.  However, $\delta_c$ can be independently 
varied by changing either $r_\zeta$ or $a_0$. From 
equation~\eqref{eq:1}, with \eqref{eq:A1} and \eqref{eq:A3}, we find 
that $\delta_c$ scales as $\sqrt{r_\zeta}$ and $a_0^{5/2}$.  With 
reference values of $r_\zeta = 5$ and $a_0 = 5$~mm, the compaction 
length is 32~km. This value is calculated at conditions representative 
of the shallow partially molten asthenosphere beneath the ridge, with  
$\phi = 0.01$, $\eta = 4\times10^{19}$~Pa-s and $\mu = 5$~Pa-s.  
The permeability under these conditions is 
$2.6\times10^{-13}$~m\textsup{2}. The values of $\delta_c=16$ and 
64~km in the parameter study above arise from $r_\zeta$ of 1.25 and 20 
respectively, and values of $\delta_c=8$ and 128~km are obtained from 
$a_0$ of 2.87 and 8.71~mm respectively.

Table~\ref{tab:par} contains a list of model parameters and their 
reference values.  Parameters not listed there are the same as in 
Table~1 of \cite{keller16}.

\begin{table}[ht]
  \centering
    \caption{\textbf{Simulation parameters, definitions, and reference values}}
  \begin{tabular}{llll}
    \hline
    Depth of domain & $D$ & km & 200 \T \\
    Width of domain & $W$ & km & 300 \\
    Mantle density & $\rho_0$ & kg/m$^{3}$ & 3200 \\
    Crustal density & $\rho_c$ & kg/m$^{3}$ & 3000 \\
    Rock to melt density contrast & $\Deltarho$ & kg/m$^{3}$ & 500 \\
    Half-spreading rate & $u_0$ & cm/yr & 3 \\
    Dry pre-exponential factor & $A_0^\mathrm{dry}$ & Pa-s/m$^3$ & 1.11\e{14} \\
    Wet pre-exponential factor & $A_0^\mathrm{wet}$ & Pa-s (10$^6$ H/Si)/m$^3$ & 6.67\e{15} \\
    Activation energy & $E_a$ & J/mol & 3.75$\times10^5$ \\
    Activation volume & $V_a$ & m$^3$/mol & 4\e{-6} \\
    Universal gas constant & $R$ & J/mol/K & 8.314 \\
    Rock viscosity melt-weakening factor & $\alpha$ & Pa-s & 30 \\
    Compaction to shear viscosity ratio & $r_\zeta$ & - & 5 \\
    Melt viscosity constant & $\mu_0$ & Pa-s & 1 \\
    Melt viscosity compositional factors & $\lambda^i$ & - & [1,10,0.1,0.01] \\
    Grain size constant & $a_0$ & m & 5\e{-3} \B \\ \hline
  \end{tabular}
  \label{tab:par}
\end{table}

\section{Mass transfer rates from simulation output 
	\label{sect:app-a}}

The following three rates of mass transfer per unit length of ridge axis are recorded to quantify melt production and focusing in simulation results. They have units of kg~yr$^{-1}$m$^{-1}$.

First, $\qb$ is the integral of component mass advected across the
base of the melting regime,
\begin{linenomath*}
  \begin{align}
    \label{eq:B1}
    \qb &= \int\limits_0^W 2 \rho_0 \phis \csi w_s~\infd x~\mid_
    {z=z_\mathrm{sol}} \: ,
  \end{align}
\end{linenomath*}
where $W$ is the width of the domain, $w_s$ is the solid upwelling
rate and $\csi$ denote the mass concentrations of chemical components
($i \in \{\text{DUN, MORB, hMORB, cMORB}\}$). The factor two
accounts for the symmetrical other half of the ridge system. The total
mantle inflow rate is obtained from the sum of component-wise rates,
$q_\mathrm{base} = \sum\limits_{i=1}^n \qb$.

Second, $\qm$ is the integral of positive component-wise melting rates
across the model domain,
\begin{linenomath*}
  \begin{align}
    \label{eq:B2}
    \qm &= \int\limits_0^W \int\limits_0^{z_\mathrm{sol}} 
    2 \Gi_+~\infd x~\infd y \: ,
  \end{align}
\end{linenomath*}
where
\begin{linenomath*}
  \begin{equation} \nonumber
    \Gi_+ =
    \begin{cases}
      \Gi & \text{if $\Gi>0$} \: \\
      0 & \text{otherwise} \:\: .
    \end{cases}
  \end{equation}
\end{linenomath*}
The total melt production rate is obtained by summing over the
components, $q_\mathrm{melt} = \sum\limits_{i=1}^n \qm$.

And third, $\qf$ is the integral of component-wise melt focused to the ridge axis,
\begin{linenomath*}
  \begin{align}
    \label{eq:B3}
    \qf &= \int_0^W 2 \rho_0 \phi \cli w_\ell ~\infd x~\mid_{z=d} \: ,
  \end{align}
\end{linenomath*}
where $d = 8$~km is the depth at the base of the imposed melt 
extraction zone in the axis and $w_\ell$ is the vertical rate of melt 
flow. The total melt focusing rate is $q_\mathrm{focus} = 
\sum\limits_{i=1}^n \qf$.

The predicted crustal thickness is calculated from the ratio of melt
focusing and plate spreading rates:
\begin{linenomath*}
  \begin{align}
    \label{eq:B4}
    H_c = q_\mathrm{focus} / (2 \rho_c u_0) \: ,
  \end{align}
\end{linenomath*}
with the density of oceanic crust $\rho_c=3000$~kg/m\textsup{3}.

Finally, extraction rates for \water{} and \carbon{} are calculated by
multiplying the focusing rates of the volatile-bearing components by
their constant volatile contents.  Hence we have
$ q^\mathrm{H_2O}_\mathrm{focus} = 0.05 \times 
q^\mathrm{hMORB}_\mathrm{focus}$, 
and
$q^\mathrm{CO_2}_\mathrm{focus} = 0.20 \times
q^\mathrm{cMORB}_\mathrm{focus}.$

\cleardoublepage

%% Supplementary material
\setcounter{page}{1}
\renewcommand{\thesection}{S} \setcounter{section}{0}
\renewcommand{\thetable}{S\arabic{table}} \setcounter{table}{0}
\renewcommand{\thefigure}{S\arabic{figure}} \setcounter{figure}{0}

\section*{Supplementary Material to: "Volatiles beneath mid-ocean ridges: deep melting, channelised transport, focusing, and metasomatism"}

by Tobias Keller, Richard F. Katz, and Marc M. Hirschmann.

\vspace{24 pt}

\begin{figure}[ppp]
  \centering
  \includegraphics[width=\textwidth]{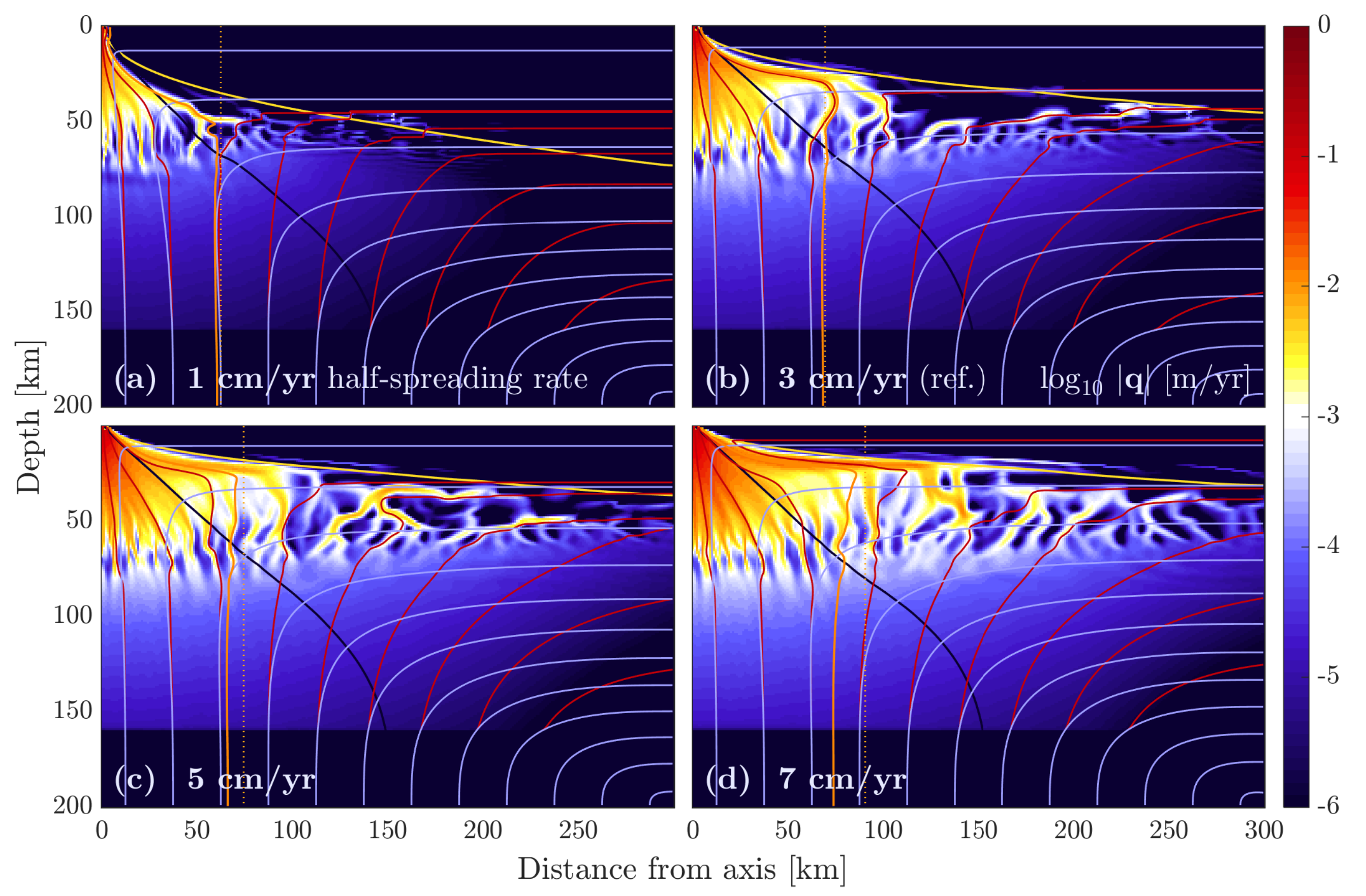}
  \caption{Darcy flux magnitude $\vert\q\vert$ at $t=10$~Myr for
    simulations with half-spreading rates of \textbf{(a)} 1~cm/yr, 
    \textbf{(b)} 3~cm/yr (reference), \textbf{(c)} 5~cm/yr, and 
    \textbf{(d)} 7~cm/yr. Streamlines for melt (red) and solid (blue) 
    flow, outline of primary upwelling domain (black), and melt 
    focusing domain (orange), with $x_e$ (dotted) for comparison.
	Faster spreading rates lead to shallower temperature structure and 
	higher melt production reflected in increased Darcy flux. The 
	focusing distance grows with spreading rate. Note that the primary 
	upwelling contours consistently intersect with the proxies for 
	melt focusing distance around the volatile-free solidus depth at $
	\sim$70 km.}
  \label{fig:S1}
\end{figure}

\begin{figure}[ppp]
  \centering
  \includegraphics[width=\textwidth]{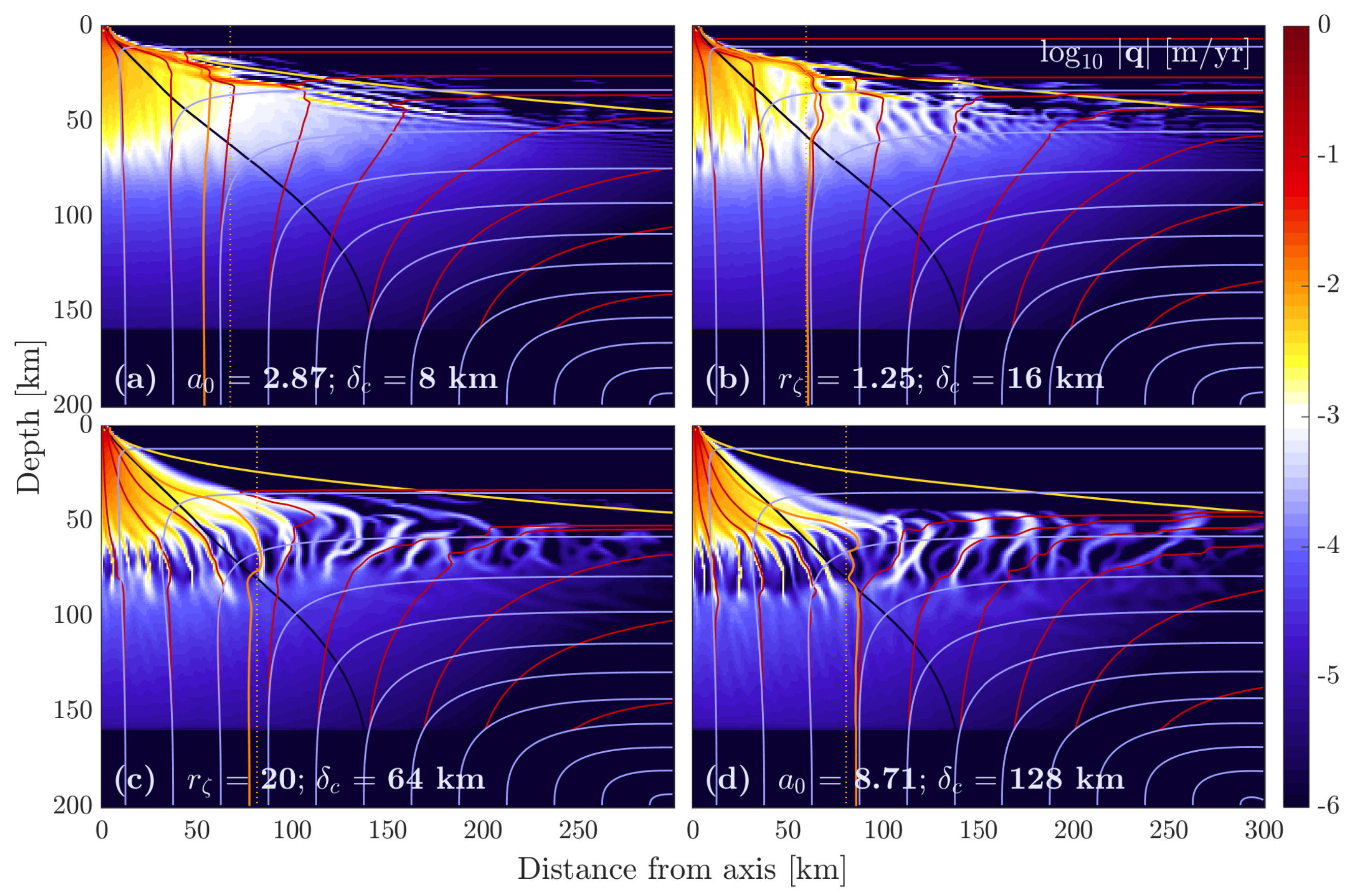}
  \caption{Darcy flux magnitude $\vert\q\vert$ at $t=10$~Myr for
    simulations with compaction lengths of \textbf{(a)} 8~km, 
    \textbf{(b)} 16~km, \textbf{(c)} 64~km, and \textbf{(d)} 128~km. 
    Streamlines for melt (red) and solid (blue) flow, outline of 
    primary upwelling domain (black), and melt focusing domain 
    (orange), with $x_e$ (dotted) for comparison.
    At small compaction length melt streamlines are dominantly 
    vertical or along inclined plane of LAB. Melt streamlines are 
    increasingly bent away from the thermal boundary layer with 
    growing compaction length. Channels are more pronounced at high 
    compaction lengths.}
  \label{fig:S2}
\end{figure}

\begin{figure}[ppp]
  \centering
  \includegraphics[width=\textwidth]{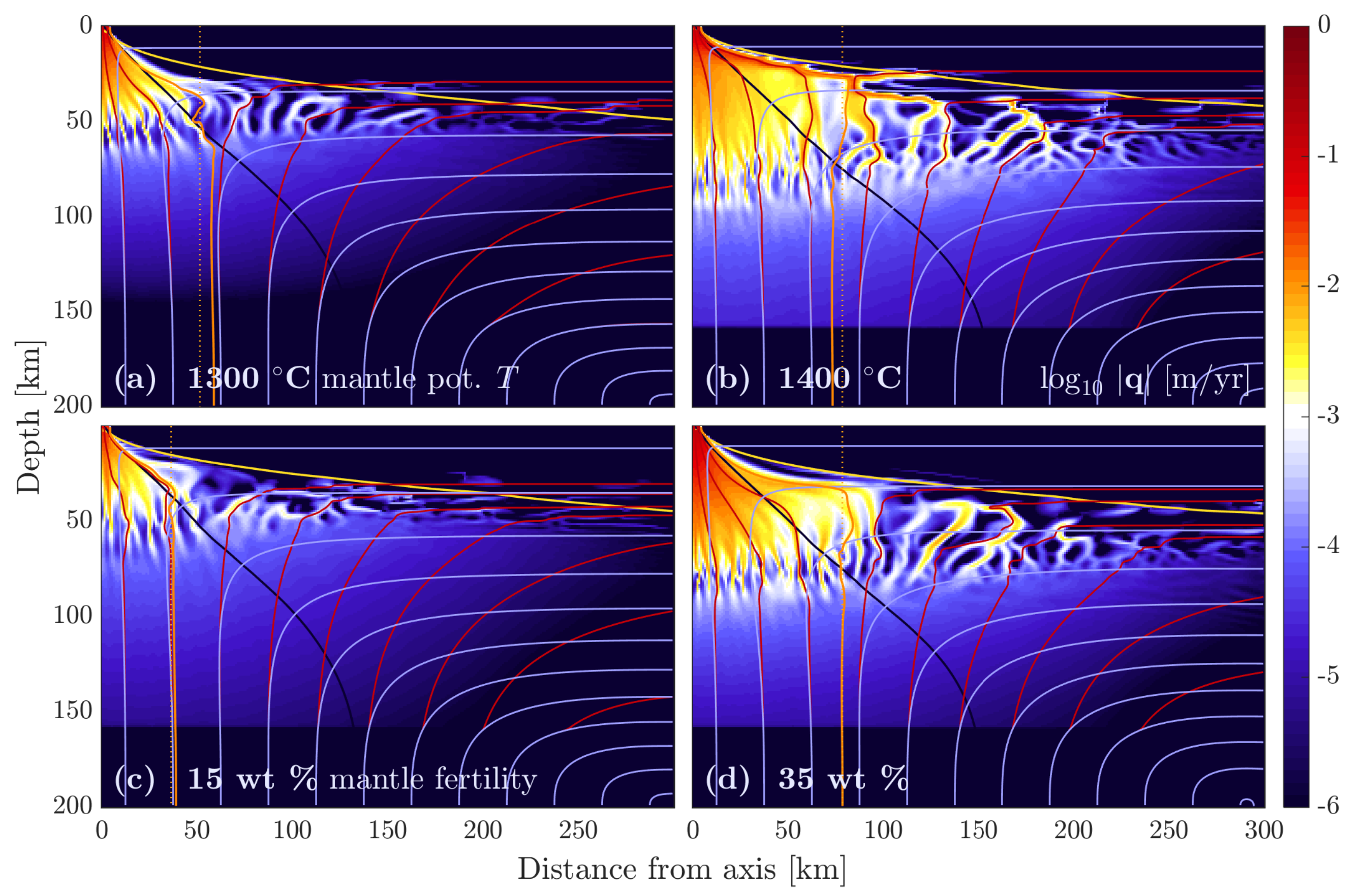}
  \caption{Darcy flux magnitude $\vert\q\vert$ at $t=10$~Myr for
    simulations with mantle potential temperatures of \textbf{(a)} 
    1300~\degC{}, and \textbf{(b)} 1400~\degC; and mantle fertilities 
    of \textbf{(c)} 15~wt\%, and \textbf{(d)} 35~wt\%. Streamlines for 
    melt (red) and solid (blue) flow, outline of primary upwelling 
    domain (black), and melt focusing domain (orange), with $x_e$ 
    (dotted) for comparison.
    Both increased temperature and fertility deepen the onset of 
    volatile-free melting and therefore lower the depth of channel 
    formation. The primary upwelling contour again intersects the 
    focusing distance proxies around that depth. The high-temperature 
    case shows signature of lower compaction length focusing regime.}
  \label{fig:S3}
\end{figure}

\begin{figure}[ppp]
  \centering
  \includegraphics[width=\textwidth]{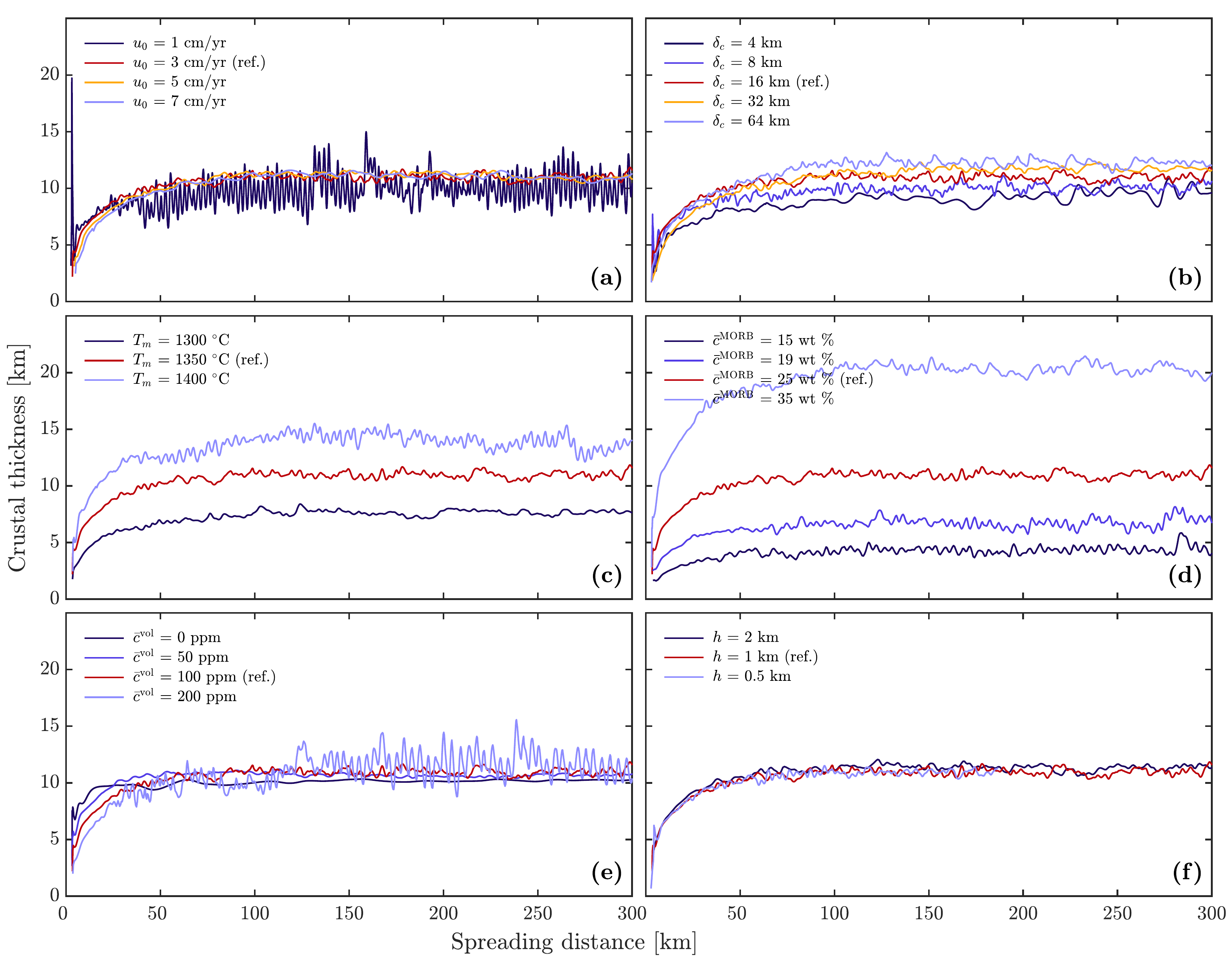}
  \caption{Time series of crustal thickness for all parameter choices. 
  	Results plotted against spreading distance. Time averages 
  	extracted over interval from 180--300~km. The resolution tests in 	
  	\textbf{h} show that crust production in the reference case is 
  	well resolved at reference conditions.}
  \label{fig:S4}
\end{figure}

\begin{figure}[ppp]
  \centering
  \includegraphics[width=\textwidth]{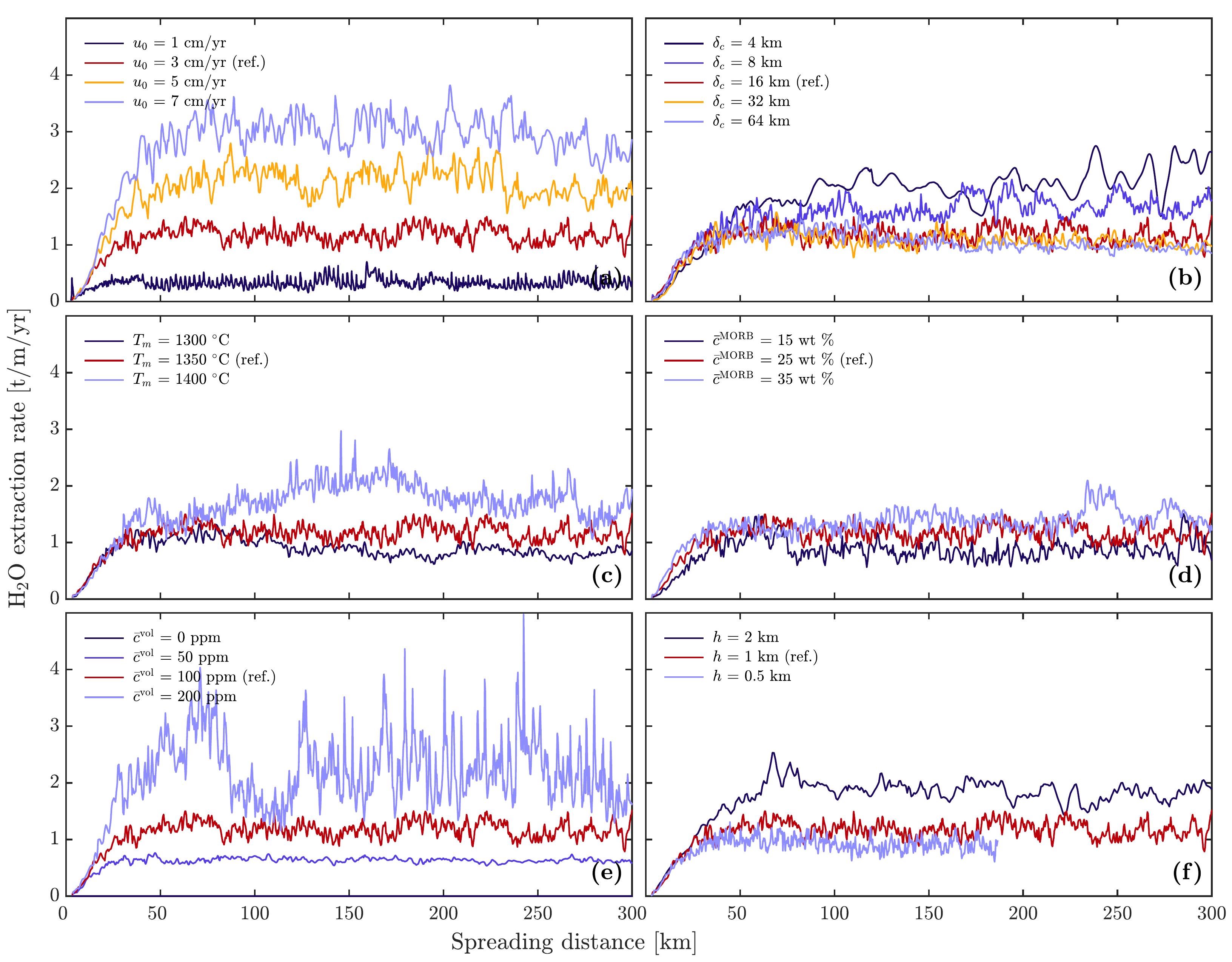}
  \caption{Time series of \water{} extraction rate for all parameter 
  	choices. Results plotted against spreading distance. Time averages 
  	extracted over interval from 180--300~km.  The resolution tests in 
  	\textbf{h} show that water extraction in the reference case is 
  	well resolved at reference conditions.}
  \label{fig:S5}
\end{figure}

\begin{figure}[ppp]
  \centering
  \includegraphics[width=\textwidth]{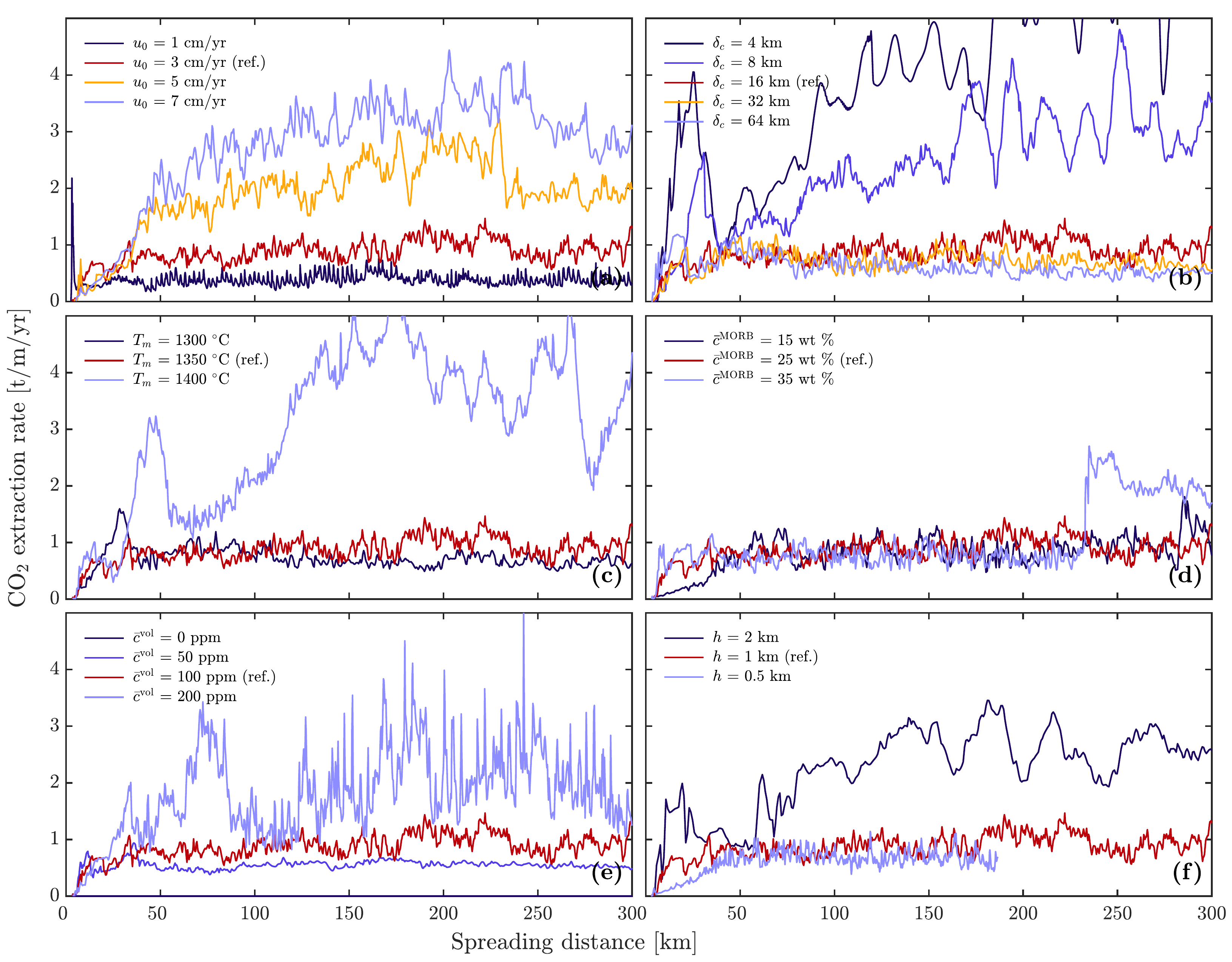}
  \caption{Time series of \carbon{} extraction rate for all parameter 
  	choices. Results plotted against spreading distance. Time averages 
  	extracted over interval from 180--300~km. The resolution tests in 	
  	\textbf{h} show that \carbon{} extraction in the reference case is 
  	well resolved at reference conditions.}
  \label{fig:S6}
\end{figure}

\begin{figure}[htb]
  \centering
  \includegraphics[width=\textwidth]{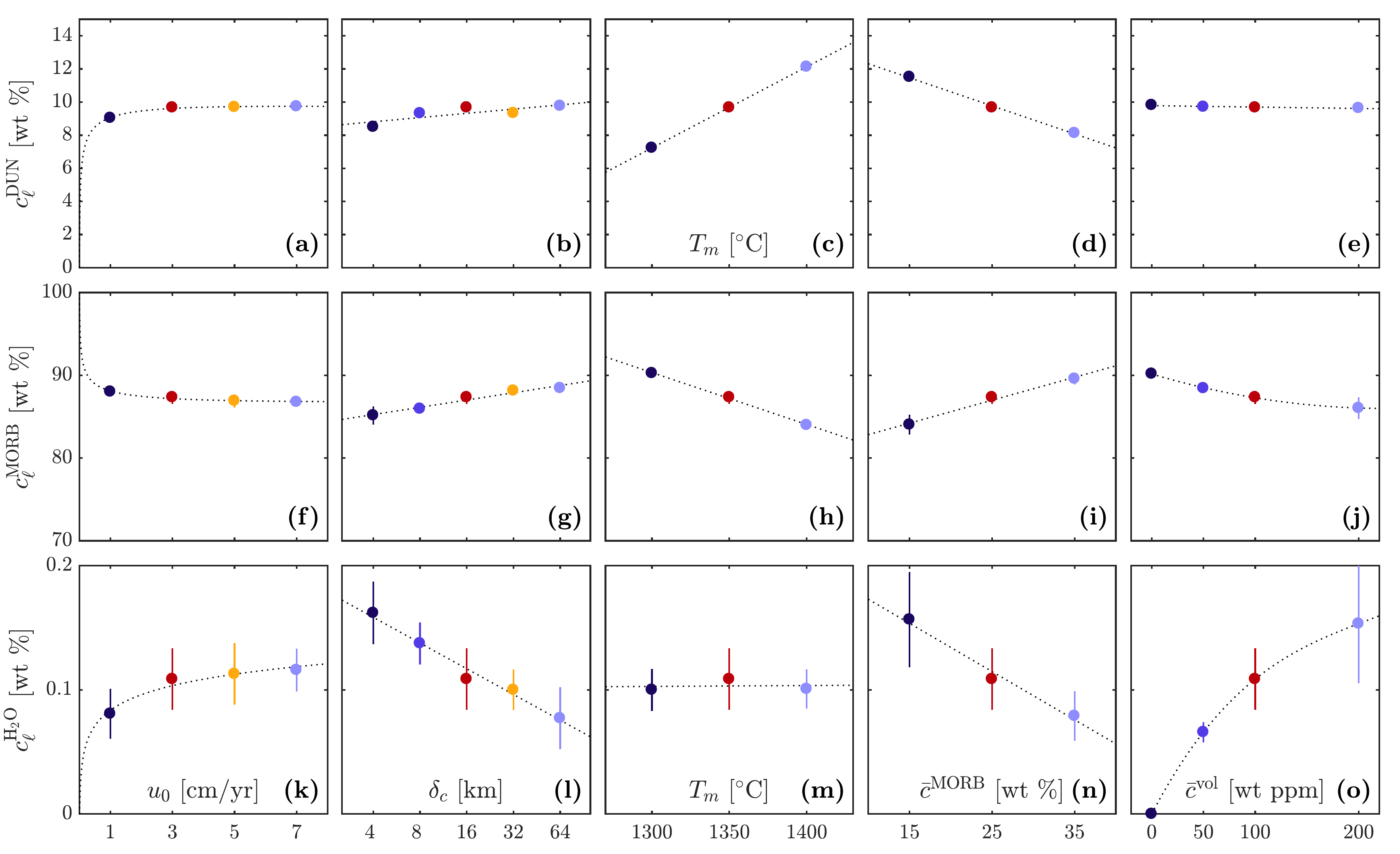}
  \caption{Component concentrations in melt focused to the ridge 
  	axis. Time averaged concentrations of dunite \textbf{(a)}--
  	\textbf{(e)}, MORB \textbf{(f)--(j)}, and \water~\textbf{(k)}--
  	\textbf{(o)}, with bars for $\pm 2 \sigma$ variability, shown as a 
  	function of: half-spreading rate, compaction length, potential 
  	temperature, fertility, and volatile content. Dotted lines are 
  	regressions of simulation outputs. }
  \label{fig:S7}
\end{figure}

The Supplementary Material below provides the Supplementary Table \ref{tab:sim} and seven Supplementary Figures, \ref{fig:S1}--\ref{fig:S7}, as referenced in the article.

\begin{landscape}
  \begin{table}[htb]
    {\centering \small
      \caption{\textbf{Parameter variations for simulation runs}}
      \label{tab:sim}
      \begin{tabular}{lccccccccccccccl}
  	Run ID & $u_0$ & $T_m$ & $c^\mathrm{morb}$ & $c^\mathrm{vol}$ & $r_\zeta$ & $a_0$ & $A_p^{morb}$ & $A_p^\mathrm{vol}$ & $\eta(c_H)$ & $\mu(c^i)$ & $\eta(\phi)$ & $D$ & $D_\mathrm{sol}^\mathrm{max}$ & $h$ & Comments \B \\\hline
               & cm/yr & \degC & wt \% & wt ppm & 1 & mm & \% & \% & - & - & - & km & km & km & units \T \B \\
	morb0 & - & - & - & - & - & - & 0 & - & - & - & - & - & - & - & high pert. ampl. \\
	\textbf{morb1} & \textbf{3} & \textbf{1350} & \textbf{25} & \textbf{100} & \textbf{5} & \textbf{5} & \textbf{10} & \textbf{0} & \textbf{on} & \textbf{on} & \textbf{on} & \textbf{200} & \textbf{160} & \textbf{1} & heterogeneous ref. \\
	morb2 & - & - & - & - & - & - & 50 & - & - & - & - & - & - & - & high pert. ampl. \\
	morb3 & 1 & - & - & - & - & - & - & - & - & - & - & - & - & - & slow-spreading \\
	morb4 & 5 & - & - & - & - & - & - & - & - & - & - & - & - & - & fast-spreading \\
	morb25 & 7 & - & - & - & - & - & - & - & - & - & - & - & - & - & fastest-spreading \\
	morb5 & - & 1300 & - & - & - & - & - & - & - & - & - & - & - & - & low pot. T \\
	morb6 & - & 1400 & - & - & - & - & - & - & - & - & - & - & - & - & high pot. T \\
	morb7 & - & - & 15 & - & - & - & - & - & - & - & - & - & - & - & low fertility \\
	morb8 & - & - & 35 & - & - & - & - & - & - & - & - & - & - & - & high fertility \\
	morb9 & - & - & - & 0 & - & - & - & - & - & - & - & - & - & - & no volatiles \\
	morb22 & - & - & - & 50 & - & - & - & - & - & - & - & - & - & - & low volatiles \\
	morb10 & - & - & - & 200 & - & - & - & - & - & - & - & - & - & - & high volatiles \\
	morb11 & - & - & - & 100 (H) & - & - & - & - & - & - & - & - & - & - & water only \\
	morb12 & - & - & - & 100 (C) & - & - & - & - & - & - & - & - & - & - & carbon only \\
	morb13 & - & - & - & - & 1.25 & - & - & - & - & - & - & - & - & - & low $r_\zeta$ \\
	morb14 & - & - & - & - & 20 & - & - & - & - & - & - & - & - & - & high $r_\zeta$ \\
	morb15 & - & - & - & - & - & 2.87 & - & - & - & - & - & - & - & - & small grain size \\
	morb16 & - & - & - & - & - & 8.71 & - & - & - & - & - & - & - & - & large grain size \\
	morb17 & - & - & - & - & - & - & - & $-$25 & - & - & - & - & - & - & anti-corr. pert. \\
	morb18 & - & - & - & - & - & - & - & 25 & - & - & - & - & - & - & corr. pert. \\
	morb23 & - & - & - & - & - & - & 0 & 25 & - & - & - & - & - & - & only vol. pert. \\
	morb19 & - & - & - & - & - & - & - & - & off & - & - & - & - & - & no H-dep. $\eta$ \\
	morb20 & - & - & - & - & - & - & - & - & - & off & - & - & - & - & no C-dep. $\mu$ \\
	morb21 & - & - & - & - & - & - & - & - & - & - & off & - & - & - & no $\phi$-dep. $\eta$ \\
	morb1d & - & - & - & - & - & - & - & - & - & - & - & 300 & - & - & domain depth test \\
	morb26 & - & - & - & - & - & - & - & - & - & - & - & - & 140 & - & shallower melting \\
	morb27 & - & - & - & - & - & - & - & - & - & - & - & - & 180 & - & deeper melting \\
	morb1l & - & - & - & - & - & - & - & - & - & - & - & - & - & 2 & low res. test \\
	morb1h & - & - & - & - & - & - & - & - & - & - & - & - & - & 0.5 & high res. test \\
		morb1s & - & - & 19 & - & - & - & - & - & - & - & - & - & - & - & shifted fertility \B \\\hline
      \end{tabular}}
  \end{table}
\end{landscape}

\pagebreak

\end{document}